\documentclass[twocolumn,bibyear]{aa}

\usepackage{lscape}  
\usepackage{graphicx}
\usepackage[varg]{txfonts}
\usepackage{natbib}
\usepackage{amsfonts}
\usepackage[colorlinks=true, linkcolor=blue, citecolor=blue, filecolor=blue, urlcolor=blue]{hyperref}
\usepackage{epsfig}
\usepackage{color}
\usepackage{xcolor}

\newcommand   {\about} {\mbox{$\sim$}}

\renewcommand {\ga}    {\mbox{\rlap{\hbox{\lower5pt\hbox{$\sim$}}}\hbox{$>$}}}
\renewcommand {\la}    {\mbox{\rlap{\hbox{\lower5pt\hbox{$\sim$}}}\hbox{$<$}}}

\begin{document}

\title{An 18 -- 25 GHz spectroscopic survey of southern hemisphere
  dense cores}

\author{Dariusz~C.~Lis\inst{1}, Karen~Willacy\inst{1},
  Liton~Majumdar\inst{2,3}, Jorge~L.~Pineda\inst{1},
  Susanna~Widicus~Weaver\inst{4,5}, and Shinji~Horiuchi\inst{6}}

\institute{Jet Propulsion Laboratory, California Institute of
  Technology, 4800 Oak Drove Drive, Pasadena, CA 91109, USA
  \and Exoplanets and Planetary Formation Group, School of Earth and
  Planetary Sciences, National Institute of Science Education and
  Research, Jatni 752050, Odisha, India 
  \and Homi Bhabha National Institute, Training School Complex,
  Anushaktinagar, Mumbai 400094, India 
  \and Department of Astronomy, University of Wisconsin-Madison, 475 N 
  Charter St, Madison, WI 53706, USA 
  \and Department of Chemistry, University of Wisconsin-Madison, 1101 
  University Ave, Madison, WI 53706, USA 
  \and CSIRO Space \& Astronomy/NASA Canberra Deep Space Communication
  Complex, PO Box 1035, Tuggeranng ACT 2901, Australia}

\date{Received 10 March 2026; accepted 7 March 2026}

\abstract{ We extended the radio K-band spectroscopic survey for
  organics in southern hemisphere dense cores by observing seven
  sources using NASA's Deep Space Network 70-m antenna in Canberra,
  Australia, over the frequency range of 18 to 25 GHz. Molecular
  column densities of NH$_3$, $c$-C$_3$H$_2$, HC$_3$N, HC$_5$N, CCS,
  C$_3$S, and $c$-C$_3$HD were derived for each source assuming LTE.
  The resulting column density ratios were compared with predictions
  of a state-of-the art astrochemical model to constrain the C/O ratio
  and chemical age of each source. Most cores have similar C/O ratios
  of $0.5 - 0.7$, much different from the best studied TMC-1 dense
  core characterized by a high C/O ratio of $\sim 1.4$. The chemical
  ages of the cores are also similar and fall between 0.6 and 5~Myr.
  The less dense cores tend to have the oldest chemical ages, as might
  be expected given that chemical timescales scale with density. Our
  results showcase the synergistic approach of combining radio
  observations using the DSS-43 antenna with state-of-the-art
  astrochemical models to study the chemical composition of southern
  hemisphere dense cores, enabling constraints on their C/O ratios and
  chemical ages, which remain largely unexplored. }


\titlerunning{Southern hemisphere dense cores}
\authorrunning{Lis et al.}

\maketitle
\nolinenumbers

\section{Introduction}
\label{sec:intro}

Recent molecular detections of aromatic compounds and a wide range of
interstellar complex organic molecules (COMs) in the Taurus Molecular
Cloud (TMC-1; see \citealt{mcguire2018,mcguire2020,cernicharo2022} and
references therein) have motivated extensive broadband spectral
surveys of this dense cloud. Two prominent examples include the Q-Band
Ultrasensitive Inspection Journey to the Obscure TMC-1 Environment
(QUIJOTE) survey conducted with the Yebes telescope
\citep{cernicharo2021}, and the GBT Observations of TMC-1: Hunting
Aromatic Molecules (GOTHAM) survey using the Green Bank Telescope
\citep{mcguire2018}. Together, these surveys have revealed an
extraordinary molecular richness of TMC-1, uncovering numerous
carbon-bearing molecules and highlighting the central role of carbon
chemistry in cold interstellar environments.

Carbon forms the backbone of organic molecules, many of which are
considered potential precursors to prebiotic species on Earth. For
this reason, carbon chemistry plays a fundamental role in chemical
pathways that may ultimately lead to habitable planetary environments
around other stars \citep{Bergin26}. The growing inventory of aromatic
species detected in TMC-1 further emphasizes this importance. Recent
discoveries include CN-substituted derivatives of aromatic molecules
such as naphthalene \citep{McGuire21}, as well as larger aromatic
species related to acenaphthylene \citep{Cernicharo24}, pyrene
\citep{Wenzel24,Wenzel25a}, and coronene \citep{Wenzel25b}. Polycyclic
aromatic hydrocarbons (PAHs) are also thought to represent a major
reservoir of interstellar carbon, with approximately 10--25\% of the
cosmic carbon budget potentially locked in these species
\citep{Tielens08}. These discoveries demonstrate that complex carbon
chemistry can proceed efficiently at the very low temperatures
characteristic of cold dense cores such as TMC-1.

Despite the remarkable chemical complexity revealed in TMC-1, it
remains unclear whether this level of molecular richness is typical of
dense cores in general or whether TMC-1 represents a particularly
favorable chemical environment. Addressing this question requires
systematic surveys of the chemical composition across a broader sample
of dense cores. Such studies are essential for identifying the
dominant chemical pathways responsible for the formation of key
carbon-bearing species, including long carbon-chain molecules and
complex organic molecules, and for determining how widespread these
chemical processes are within interstellar molecular clouds.

\begin{table*}
\begin{center}  
\caption{Coordinates, sizes, and temperatures of the observed cores.} 
\label{tab:param}
\begin{tabular}{lcccc}
\hline \hline 
  \rule[-3mm]{0mm}{8mm}
  Source & Position & Reference & Distance & Time \\
  \hline %\rule{0pt}{2ex}
  HOPS-108 & 05:35:27.08 --05:10:00.1 & 05:35:51.23 --05:10:00.1 &
    0.4 & 28.5 \\
  HOPS-373 & 05:46:30.91 --00:02:35.2 &	05:46:42.52 --00:07:50.6 &
    0.4 & 18.4 \\
  Vela C IRS 31 14 & 08:58:11.60 --42:37:33.6 & 08:57:48.00 --42:33:25.0 &
    0.9 & 18.7 \\
  BHR71-IRS1 & 12:01:36.50 --65:08:49.4 & 12:00:58.07 --65:10:46.0 &
   0.2 & 1.8 \\
  \hline
  G300.91+0.88 & 12:34:14.26 --61:55:22.7 & 12:33:51.50 --61:59:35.0 &
   4.4 & 18.3 \\ 
  G304.76+1.34 & 13:06:45.73 --61:28:39.9 & 13:07:31.40 --61:31:00.0 &
   -- & 6.8 \\ 
  G309.91+0.32 & 13:50:54.57 --61:44:22.1 & 13:50:25.00 --61:47:50.0 &
   5.5 & 4.1 \\
  \hline
\end{tabular}
\end{center}
Notes: Entries in the table are source, ON and OFF positions (J2000), distance (kpc),
and on-source observing time (hr).
\end{table*}

\cite{lis2025} presented radio K-band (18--25~GHz) observations of two
cores in the southern Chamaeleon complex carried out using NASA's Deep
Space Network 70-m antenna in Canberra, Australia (DSS-43), covering
the frequency range of 18--25~GHz. They surveyed the Class~0 protostar
Cha-MMS1 and the prestellar core Cha-C2, which represent different
evolutionary stages of dense cores. Several molecules were detected,
including the carbon-chain species HC$_3$N, HC$_5$N, C$_4$H, CCS, C$_3$S,
$c$-C$_3$H$_2$, $c$-C$_3$HD, and NH$_3$. A longer cyanopolyyne, HC$_7$N, was
also detected with high confidence through spectral stacking analysis.
While the molecular column densities in the two Chamaeleon cores are
typically an order of magnitude lower than those at the cyanopolyyne
peak in TMC-1, the molecular abundance ratios are generally consistent
with the values observed in TMC-1. The two exceptions are $c$-C$_3$H$_2$,
which is enhanced by a factor of \about 25 relative to the cyanopolyynes
in the Chamaeleon cores, and NH$_3$, which is enhanced by a factor of
\about 125.

Here, we extend the DSS-43 radio K-band observations to seven
additional southern hemisphere dense cores and compare their molecular
abundance ratios to state-of-the-art time dependent chemical models.
In Sect.~\ref{sec:dsn}, we discuss the DSS-43 observations, in
Sect.~\ref{sec:properties} we derive the properties of the cores and
molecular column densities, and in Sect.~\ref{sec:chemmod} we compare
their chemical composition with model predictions. In
Sect.~\ref{sec:conclusion}, we summarize our results and describe the
next steps needed to trace the evolution of prebiotic compounds in
dense cores.

\section{Deep Space Network observations}
\label{sec:dsn}

The DSN technical capabilities for astrochemistry research have
improved significantly in recent years with the installation of a new
cryogenic dual-horn dual-polarization 17–27 GHz receiver at the DSS-43
in Canberra, Australia \citep{kuiper2019}, equipped with a broadband
digital spectrometer \citep{virkler2020} that has 8 GHz of
instantaneous bandwidth, allowing for observations of multiple
transitions of heavy molecular species. Such observations can be used
to characterize the density and temperature of the gas with radiative
transfer models and derive dynamical information from the line width
and velocity at the line peak. The spectrometer provides sufficiently
high spectral resolution (30.5 kHz, or $0.35 -0.49$ km\,s$^{-1}$) to
resolve narrow molecular line shapes even in the coldest regions
($\sim 10$~K) of dense cores (see \citealt{lis2025}).

A list of molecular transitions covered by the DSS-43 K-Band receiver
is given in Table~1 of \cite{lis2025} and complete molecular
spectroscopy data can be obtained from the Cologne Database for
Molecular Spectroscopy (CDMS) catalog
\cite{muller2001,muller2005}.\footnote{Accesible online at
  https://cdms.astro.uni-koeln.de/classic/, from the CLASS Weeds
  package, or via the Splatalog interface at
  https://splatalogue.online/.}

The source sample for the present observations includes three sources
from the ALMA COMPASS program (ALMA Large Program 2022.1.00316L):
HOPS-108, HOPS373, BHR-71. In addition, we observed Vela C IRS 31 12,
and four sources from the deuterated ammonia survey of
\cite{wienen2021} characterized by low kinetic temperatures and high
D/H ratios in ammonia: G300.91+0.88, G304.76+1.34, and G309.91+0.32.
All targets are easily accessible to DSS-43, without conflicts with
deep-space communications, which is the primary objective of the
antenna. The source coordinates, including the reference positions
used for the DSS-43 position-switched observations, are listed in
Table~\ref{tab:param}. The reference positions were chosen as
emission-free regions in the \emph{Herschel}/SPIRE 350~$\mu$m dust
continuum images downloaded from the ESA \emph{Herschel} Science
Archive\footnote{https://archives.esac.esa.int/hsa/whsa/}. The SPIRE
images were also used to constrain the source size needed for
determination of molecular column densities (Sect. \ref{sec:cden}).

All of the targeted sources are embedded protostars or prestellar
cores. HOPS-108 and HOPS 373 are compact protostars identified as part
of the Herschel Orion Protostar Survey \citep{furlan2016} and also
characterized by VLA and ALMA observations as having rich chemistry
\citep{tobin2019}. Vela C IRS 31 12 is a class I protostar also
displaying rich chemistry as observed in 3 mm observations using Mopra
\citep{saul2022}. The sources G300.91+0.88, G304.76+1.34, and
G309.91+0.32 were all identified as embedded protostellar cores in the
APEX Telescope Large Area Survey of the Galaxy (ATLAS GAL) program
\citep{urquhart2017,urquhart2019} and were specifically targeted here
because of the previous detections of deuterated ammonia
\citep{wienen2021}.

Several DSS-43 observing runs were carried out between 2024 April and
2025 March. The DSN Canberra K-band digital spectrometer processes
sixteen 1-GHz wide bands, which are split into eight separate bands
from 18 to 26 GHz for each polarization \citep{virkler2020}. Each band
consists of 32,768 channels with a 30.5 kHz resolution, corresponding
to a velocity resolution of $0.35 - 0.49$~km\,s$^{-1}$, depending on
the frequency. Typical system temperatures at the elevation of the
sources were \about 53 -- 63~K. The total on-source integration times
varied between 1.8 and 28 hours (Table~\ref{tab:param}).

The raw data from the spectrometer were processed into calibrated
ON-OFF spectra using the standard DSS-43 data reduction pipeline. The
system temperature, continuously monitored using a power meter and
scaled with a factor derived using a noise diode and an ambient load
before the observation, was used in the standard ON-OFF calibration to
obtain spectra in antenna temperature units, $T_A^*$. We refer the
reader to \cite{kuiper2019} for details on the absolute system and
receiver temperature calibration. A relative gain correction was
applied to the data to account for antenna deformations as a function
of elevation.

The beam efficiency was not measured directly during our observations.
To convert the observed spectral intensities to the main beam
brightness temperature units, we used the main beam efficiency of
$\eta_{mb}=50$\% \citep{pineda2019,lis2025} rather than the measured
DSN aperture efficiency of $\eta_A = 35.5$\%. This choice is justified
given the expected extent of the molecular emission. Moreover, the
absolute intensity calibration is not critical, as opacity effects are
small or moderate for all detected lines except for ammonia. We also
used only the relative abundance ratios among the molecules rather
than absolute abundances with respect to H$_2$ in the comparison with
TMC-1.

Subsequent data reduction was carried out using the IRAM CLASS data
reduction software.\footnote{https://www.iram.fr/IRAMFR/GILDAS/} The
data reduction included blanking of noisy channels near the band edges
and removing linear baselines from individual scans fit in the
vicinity of each line of interest. The resulting baseline-removed
spectra were then averaged with 1/$\sigma^2$ weighting to produce the
final spectra used in the analysis.

\begin{figure}
\centering

\includegraphics[trim=3.2cm 2.8cm 2.5cm 5cm, clip=true,
width=0.95\columnwidth,angle=0]{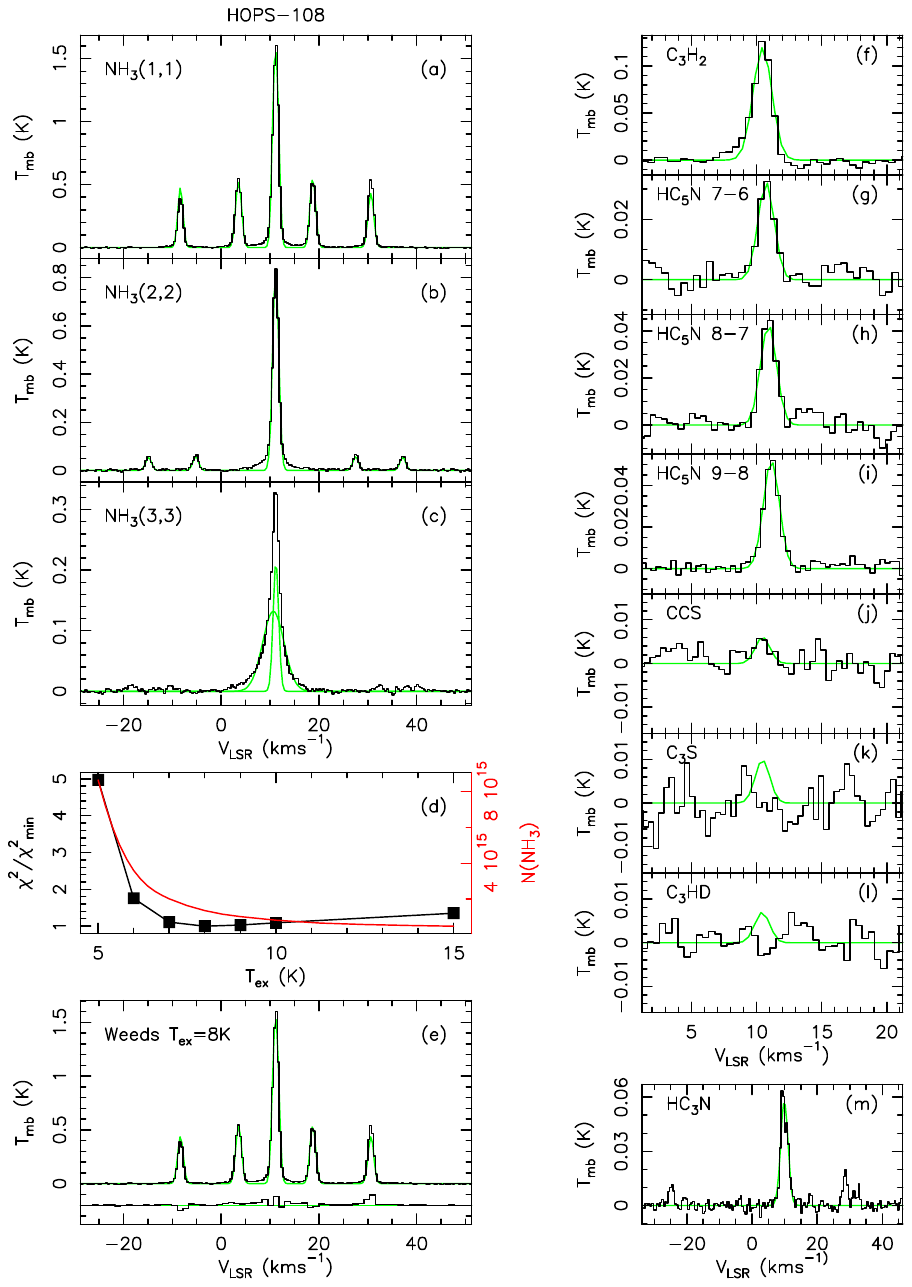} %L B R T 

\caption{DSS-43 observations of HOPS-108. (Panels a--c) Ammonia
  inversion lines. Green curves correspond to HFS fits to the (1,1)
  and (2,2) transitions and a two-component gaussian fit to the (3,3)
  transition. (Panel d) Normalized $\chi^2$ of the Weeds fit as a
  function of the excitation temperature. (Panel e) Best Weeds fit to
  the (1,1) transition corresponding to an excitation temperature of
  8~K. (Panels f-m) Spectra of other transitions included in the
  analysis. Green lines correspond to a single component Gaussian fit
  and 3$\sigma$ upper limits for CCS, C$_3$S, and $c$-C$_3$HD. }
\label{fig:hops108}
\end{figure}

The resulting spectra of HOPS-108 are shown in
Figure~\ref{fig:hops108}, as an example. Molecules included in our
analysis are NH$_3$, $c$-C$_3$H$_2$, HC$_3$N, HC$_5$N, CCS,
C$_3$S, and $c$-C$_3$HD. Given the variations in the observing time among
sources and the intrinsic line strengths, not all lines are detected
in all sources. In cases of non-detections, we report 3$\sigma$ upper
limits based on the noise measured in the spectra at the corresponding
line frequencies. In the case of HOPS-108, upper limits are determined
for CCS, C$_3$S, and $c$-C$_3$HD (Fig.~\ref{fig:hops108}, panels j--l).
Spectra of the remaining sources are shown in
the Appendix.%~\ref{app:spectra}.

\section{Results}
\label{sec:properties}

Using the spectral fits to the DSS-43 spectra, we first derived the
gas kinetic and excitation temperatures and estimated H$_2$ densities
based on observations of ammonia inversion lines. We then determined
molecular column densities using the Weeds modeling package\footnote{
  https://www.iram.fr/IRAMFR/GILDAS/doc/html/weeds-html/ weeds.html},
which can perform simple modeling of the observed spectra under the
assumption of local thermodynamical equilibrium (LTE). The LTE
approximation is justified as collisional cross-sections are not
available for some of the species considered here.

\subsection{Gas temperatures}
\label{sec:temp}

Calculations of molecular column densities require prior knowledge of
the excitation temperature (via LTE calculations) or kinetic
temperature and density (via radiative transfer models, such as the
large velocity gradient, LVG, approach). The kinetic temperature can
be estimated from observations of the ammonia inversion lines (see
Appendix B of \citealt{lis2025}). The ammonia (1,1) to (3,3) inversion
lines are detected with high signal-to-noise ratio in all sources. In
all sources, the (1,1) and (2,2) lines are characterized by narrow
line widths, with the hyperfine structure clearly detected
(Figure~\ref{fig:hops108}, panels a--b). However, the (3,3) line in
some sources shows an additional broader component originating in the
outflow driven by an embedded protostellar source (see
Figure~\ref{fig:hops108}, panel c, for an example). The broad
component is most prominent in the (3,3) line, as this high-energy
transition is only weakly excited in the cold gas responsible for the
narrow component emission. In such cases, we fitted the line
profile with two Gaussian components to separate the narrow-line
emission for temperature determination.

The resulting estimates of the kinetic temperature for all DSS-43
targets based on the observed line intensities of the ammonia
inversion lines are listed in Table~\ref{tab:cden}. They are in 
agreement with the values reported by \cite{wienen2018}.

To determine the ammonia excitation temperature, we performed
least-squares fits to the ammonia (1,1) spectra for different
excitation temperatures with the ammonia column density, line center
velocity, and line width as free parameters. The source size was fixed
at the value derived from the SPIRE 350~$\mu$m observations.
Figure~\ref{fig:hops108}, panel (d) shows the normalized $\chi^2$ as a
function of the excitation temperature for HOPS-108. The minimum is
reached for $T_{ex}=8$~K, as compared to the kinetic temperature of
19.6~K. The Weeds best fit to the ammonia (1,1) spectrum for this
excitation temperature is shown as the green line in
Figure~\ref{fig:hops108}, panel (e), with the residuals shown below. 
Fits for other sources are shown in the Appendix.%~\ref{app:spectra}.
The resulting ammonia excitation temperatures for all DSS-43 targets
are listed in Table~\ref{tab:cden}.

Although the fitting procedure used here for the determination of the
ammonia excitation temperature is different from the simple hyperfine
structure fit used by \cite{lis2025}, we confirmed that it gives 
consistent results for the Chamaeleon cores. In the case of Cha-C2, 
we derive the same excitation temperature of 5.5~K using the method 
presented here, while for Cha-MMS1, our best fit excitation 
temperature using Weeds is 7.0~K compared to 7.6~K in
\cite{lis2025}. We note that the ammonia column density is sensitive
to $T_{ex}$ at low excitation temperatures (see red line in
Figure~\ref{fig:hops108}, panel d). In the case of Cha-MMS1, lowering
the excitation temperature from 7.6 to 7~K increases the column
density by $\sim$25\%.

\subsection{Gas densities}
\label{sec:dens}

Since collisional rate coefficients are available for ammonia, the
combination of the kinetic and excitation temperature can be used to
estimate the H$_2$ density using LVG models. We used the offline
version of the RADEX code \citep{vandertak2007} to calculate the
ammonia excitation temperature as a function of density for a given
kinetic temperature. Several sets of collisional cross-sections with
$o-$ and $p-$H$_2$ are available in the literature
\citep{danby1988,demes2023,loreau2023}. The \cite{demes2023}
calculations are for collisions with both $o-$ and $p-$H$_2$. The
\cite{danby1988} and \cite{loreau2023} cross-sections are only for
collisions with $p-$H$_2$.  However, the \cite{loreau2023} calculations
resolve the NH$_3$ hyperfine structure.

\begin{figure}
\centering

\includegraphics[trim=2cm 2.5cm 5.9cm 2.5cm, clip=true,  
width=0.95\columnwidth,angle=0]{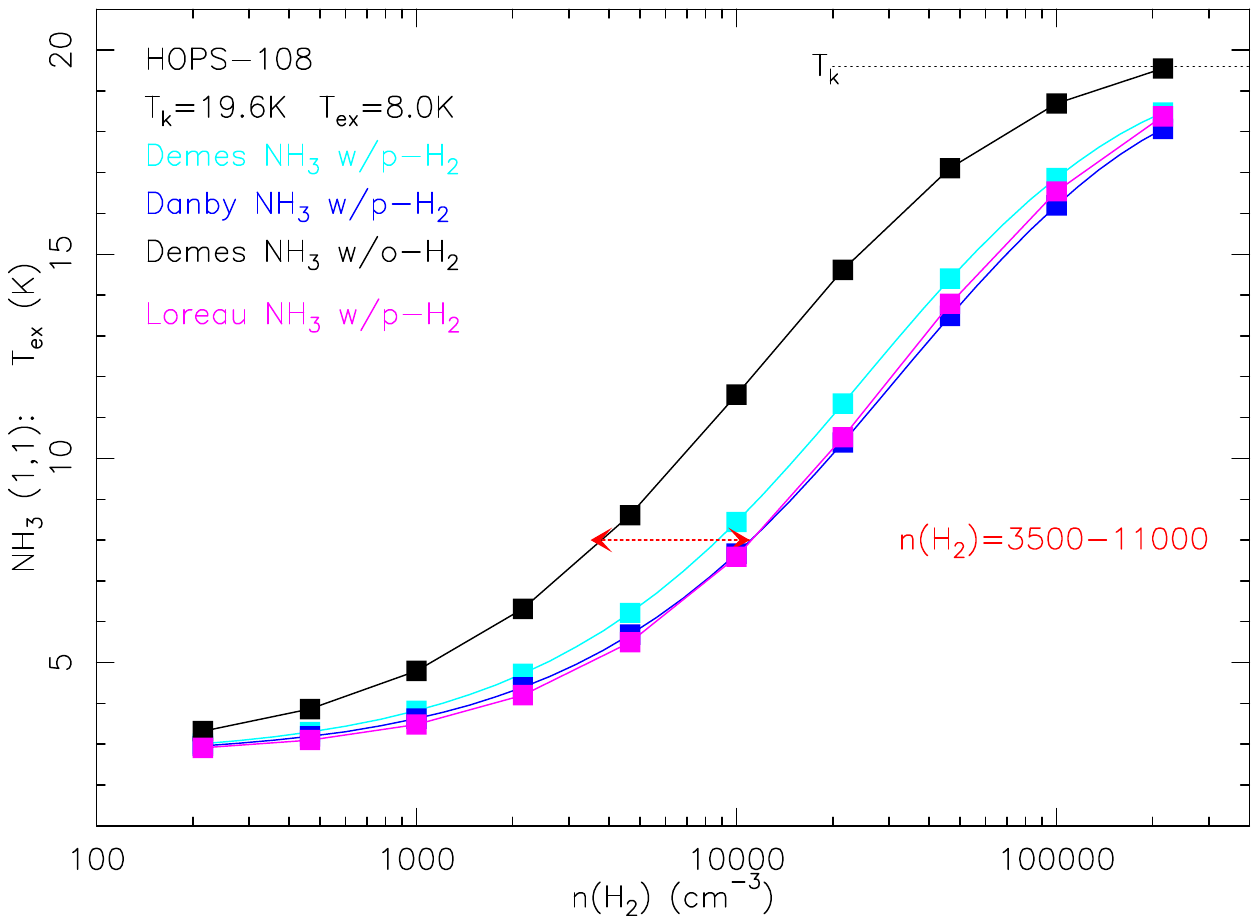} %L B R T  
\includegraphics[trim=2cm 2.5cm 5.9cm 2.5cm, clip=true,  
width=0.95\columnwidth,angle=0]{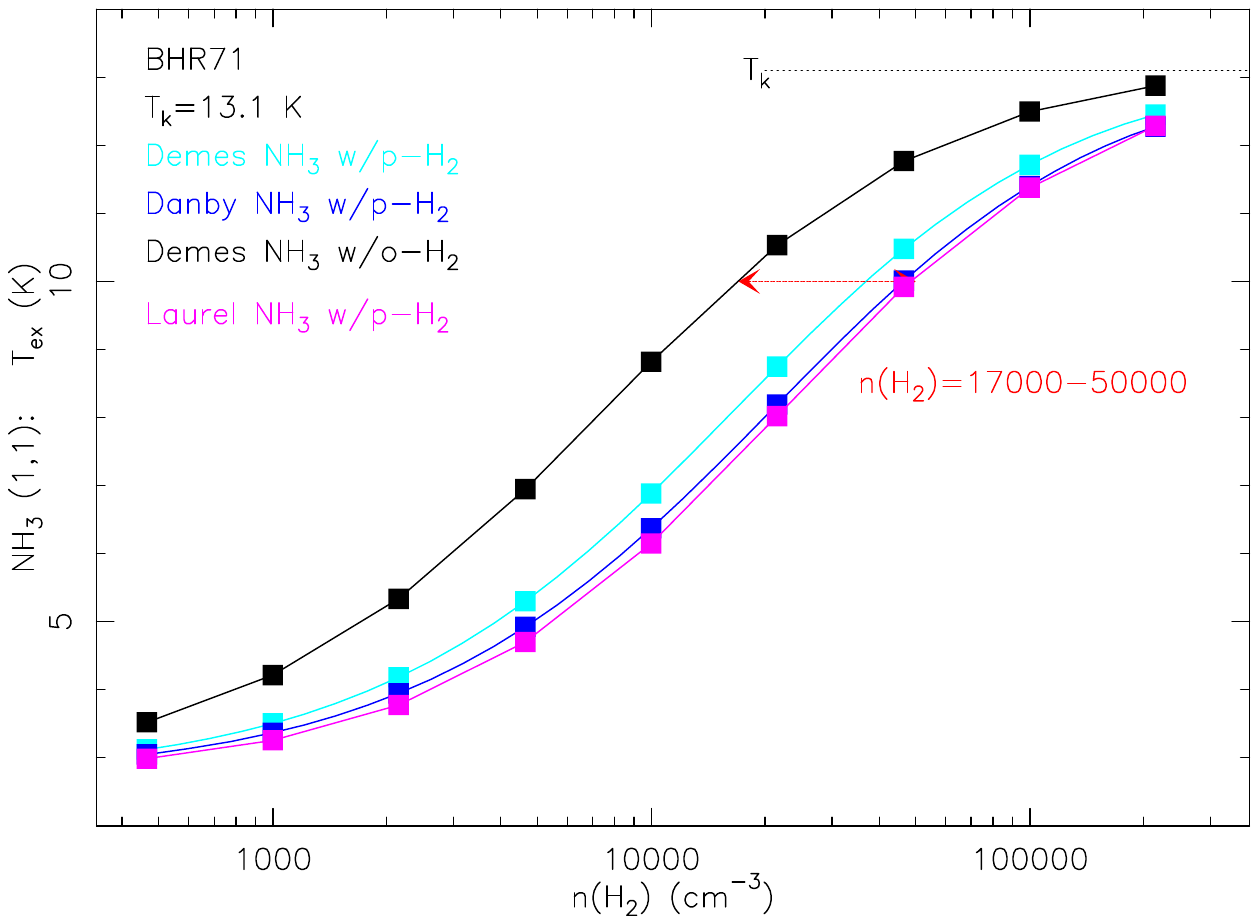} %L B R T
\includegraphics[trim=2cm 2.5cm 5.9cm 2.5cm, clip=true,  
width=0.95\columnwidth,angle=0]{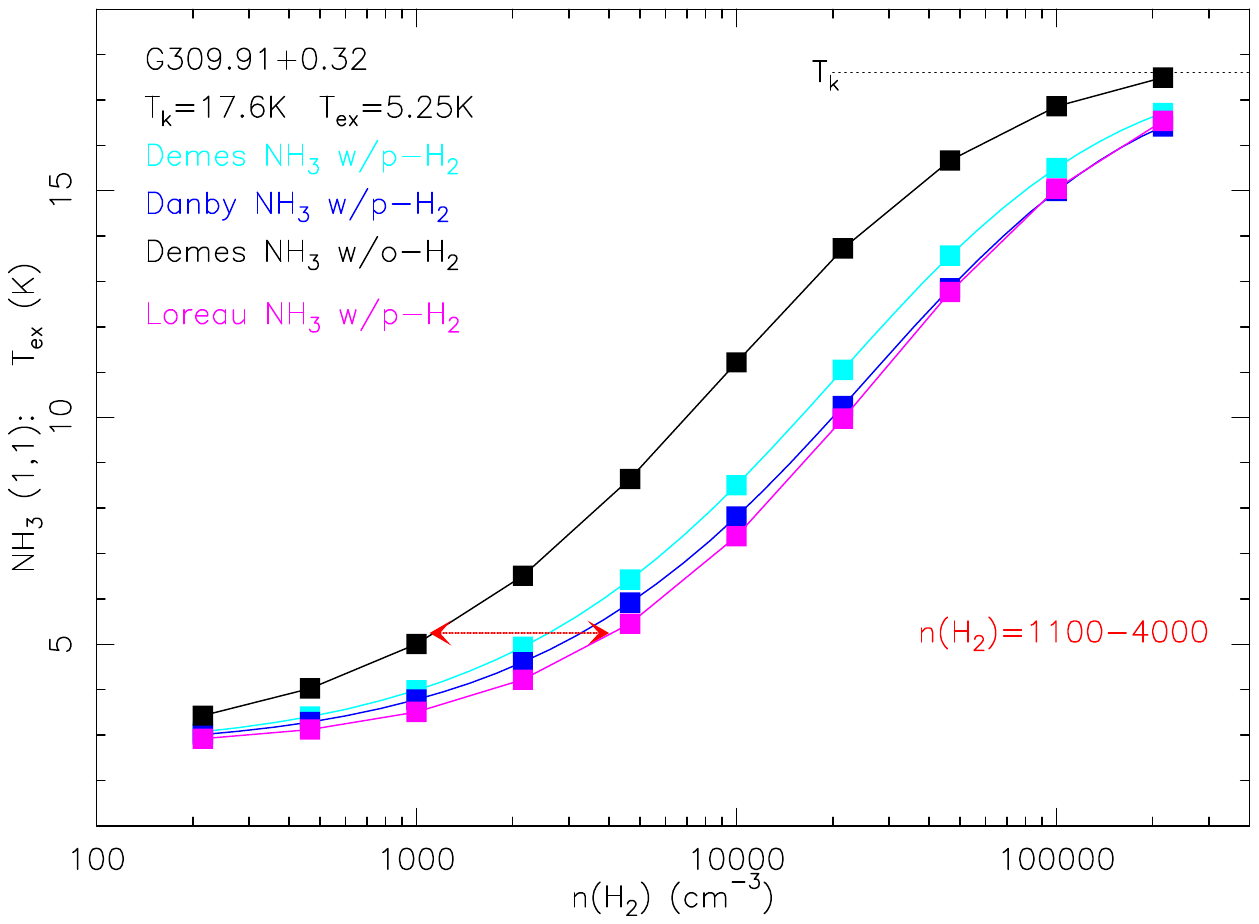} %L B R T  
\caption{LVG models of the NH$_3$ (1,1) excitation in HOPS-108,
  BHR71-IRS1, and G309.91+0.32, top to bottom panels, respectively. The
  color curves correspond to different collisional cross-sections, as
  described in the legend.}
\label{fig:tex}
\end{figure}

Figure~\ref{fig:tex} (top panel) shows the excitation temperature of the NH$_3$
(1,1) line (the strongest HFS component in the case of
\citealt{loreau2023} cross-sections) in HOPS-108 as a function of the
gas density. The calculations assume a kinetic temperature of 19.6~K,
as determined above, a FWHM line width of 1.3~km\,s$^{-1}$, and an
NH$_3$ column density of $1.5 \times 10^{14}$ cm$^{-2}$, which
approximately reproduces the observed intensity of the main HFS
component with the \cite{loreau2023} cross-sections.\footnote{Note
  that the column densities used in the LVG calculations are averaged
  values in the DSN beam, while those reported in Table~\ref{tab:cden}
  are peak values in a pencil beam.} The color curves correspond to
different collisional cross-sections. The three sets of collisional
cross-sections with $p$-H$_2$ (cyan, blue, and magenta) are in good
agreement. For a given density, collisions with $o$-H$_2$ result in a
higher excitation temperature, as $o-$H$_2$ is more efficient at
collisional excitation than $p-$H$_2$. The excitation temperature of
8~K derived for HOPS-108 from the Weeds models ($T_{ex}/T_k = 0.41$) 
corresponds to a range of densities of $3.5 \times 10^3 - 1.1 \times 
10^4$~cm$^{-3}$, depending on the collisional cross-section set used.

We carried out similar calculations for BHR71-IRS1
(Figure~\ref{fig:tex}, middle panel), which has the
highest $T_{ex}/T_k = 0.76$ in our sample, assuming a column density
of $8.0 \times 10^{13}$ cm$^{-2}$. The resulting density range is
$1.7 \times 10^4 - 5.0 \times 10^4$~cm$^{-3}$. G309.91+0.32
(Figure~\ref{fig:tex}, bottom panel) is
characterized by one of the lowest $T_{ex}/T_k = 0.30$ in our sample.
In this case, assuming an NH$_3$
column density of $3.8 \times 10^{14}$ cm$^{-2}$, the resulting
density range is $1.1 \times 10^3 - 4.0 \times 10^3$~cm$^{-3}$. 

In general, the most distant sources have lower volume densities, as
the emission in the DSS-43 beam is dominated by extended, lower
density envelopes. BHR71-IRS1, being the closest and relatively
isolated source, has the highest volume density. These calculations
are not meant to provide accurate density estimates in our target
sources, but instead to constrain a relevant density range for the
chemical models. Based on the ratio of the excitation to kinetic
temperature, we expect Cha-MMS1 to have similar density to BHR71-IRS1.
Vela C IRS 21 14, HOPS-373, and Cha-C2 should be similar to HOPS-108,
while G300.91+0.88 and G304.76+1.34 should be similar to G309.01+0.32.
We note that the densities reported here are average values in the
relatively large DSS-43 beam. \emph{Herschel} dust continuum images
indicate the presence of compact sources, which will be characterized
by higher densities.

\begin{table*}
\begin{center}  
\caption{Source sizes, temperatures, and molecular column densities.} 
\label{tab:cden}
\begin{tabular}{lcccccccccc}
\hline \hline 
  \rule[-3mm]{0mm}{8mm}
  Source & Size & $T_k$ & $T_{ex}$ & NH$_3$ & $c$-C$_3$H$_2$ & HC$_3$N &
     HC$_5$N & CCS & C$_3$S & $c$-C$_3$HD \\
  \hline %\rule{0pt}{2ex}
  HOPS-108      & 58 & 19.6 & 8.0 & 1.36(15) & 1.05(13) & 9.71(12)
  & 1.95(12) &     5.27(12)$^a$ & 3.57(11)$^a$ & 8.19(11)$^a$ \\
  HOPS-373      & 50 & 14.9 & 7.0 & 6.63(14) & 8.88(12) & 1.00(12)$^a$
  & 2.56(11)$^a$ & 1.13(12) & 4.22(11)$^a$ & 1.12(12)$^a$ \\
  Vela C IRS 31 14 & 41 & 15.1 & 6.0 & 2.49(15) & 1.83(13) & 6.60(12)
  & 1.57(12) & 3.77(12) & 7.93(11) &1.38(12)$^a$\\
  BHR71-IRS1    & 24 & 13.1 & 10.0 & 1.77(15) & 1.22(14) & 3.81(13)
  & 7.90(12) & 9.78212) & 7.54(12)$^b$ & 1.31(13)$^a$\\
  \hline
  G300.91+0.88  & 68 & 14.7 & 4.5 & \textit{1.76(16)} & 2.21(13) & 8.15(12)
  & 1.52(12) & 4.49(12) & 1.46(12) & 7.95(11)$^a$\\
  G304.76+1.34  & 109  & 13.7 & 4.3 & \textit{1.40(16)} & 1.73(13) & 4.80(12) 
  & 1.26(12) & 3.99(12) & 9.71(11) & 1.09(12)$^a$\\
  G309.91+0.32  & 66 & 17.6 & 5.2 & \textit{9.82(15)} & 1.29(13) & 7.34(12)
  & 1.53(12) & 2.72(12) & 9.70(11)$^a$ & 2.08(12)\\
  \hline
  G300.91+0.88  & 68 & 14.7 & 7.0 & 2.13(15) & 2.01(13) & 7.53(12)
  & 1.12(12) & 4.70(12) & 1.22(12) & 8.08(11)$^a$\\
  G304.76+1.34  & 109  & 13.7 & 7.0 & 1.32(15) & 1.56(13) & 3.06(12) 
  & 7.98(11) & 4.04(12) & 7.71(11) & 1.07(12)$^a$\\
  G309.91+0.32  & 66 & 17.6 & 7.0 & 2.86(15) & 1.35(13) & 7.34(12)
  & 1.35(12) & 3.00(12) & 9.11(11)$^a$ & 2.23(12)\\
  \hline
\end{tabular}
\end{center}
Notes: Entries in the table are source, source size (arcsec), kinetic
and excitation temperatures (K), and molecular column densities
(cm$^{-2}$, with the value in parenthesis being the exponent).
$^a$3$\sigma$ upper limit. $^b$Given the difference in the center
velocity and line width compared to CCS, we consider this an upper
limit. The bottom three lines show revised column densities
G300.91+0.88, G304.76+1.34, and G309.91+0.32 assuming an excitation
temperature of 7.0~K, as described in the text.
\end{table*}

\begin{table*}
\begin{center}  
\caption{Molecular column density ratios.} 
\label{tab:crat}
\begin{tabular}{lcccccccccc}
\hline \hline 
  \rule[-3mm]{0mm}{8mm}
  Source & NH$_3$/HC$_5$N & NH$_3$/$c$-C$_3$H$_2$ & $c$-C$_3$H$_2$/HC$_5$N & $c$-C$_3$HD/$c$-C$_3$H$_2$ & HC$_3$N/HC$_5$N & CCS/HC$_5$N	& CCS/C$_3$S \\ 
  \hline %\rule{0pt}{2ex}
  HOPS--108      &  700   & 130 & 5.4      & 0.08$^a$ & 5.0 & 0.27$^a$ & -- \\
  HOPS--373      &  -- & 75  &  --  & 0.13$^a$ & -- & -- & --\\
  Vela C IRS 31 14 & 1,600 & 140 & 12     & 0.08$^a$ & 4.2 & 2.4      & 4.8 \\
  BHR71--IRS1    &  220   & 15  & 15       & 0.11$^a$ & 4.8 & 1.2      & 1.3 \\
  \hline
  G300.91+0.88  & \textit{12,000} & \textit{830} & 14       & 0.04$^a$ & 5.4 & 3.0      & 3.1 \\
  G304.76+1.34  & \textit{11,000} & \textit{810} & 14       & 0.06$^a$ & 4.0 & 3.2      & 4.1 \\
  G309.91+0.32  & \textit{6,200}  & \textit{760} & 8.2      & 0.16$^a$ & 4.6 & 1.8      & 2.9 \\
  \hline
  Cha MMS1      & 1000  & 120 & 8.3       & 0.22     & 4.6 & 3.8      & 5.6 \\
  Cha C2        & 1200  & 170 & 7.1       & 0.24     & 3.7 & 6.8      & 5.4 \\
  TMC-1         & 8.5   & 26 & 0.32       & 0.28     & 3.9 & 1.7      & 7.1 \\
  \hline
  G300.91+0.88  & 1,890 & 105 & 18       & 0.04$^a$ & 6.7 & 4.2      & 3.9 \\
  G304.76+1.34  & 1,650 &  85 & 20       & 0.07$^a$ & 5.6 & 5.1      & 5.2\\
  G309.91+0.32  & 1,210 & 210 & 10     & 0.17$^a$ & 5.7 & 2.3      & 3.3 \\
  \hline
\end{tabular}
\end{center}
Notes: Entries in the table are source and selected molecular column 
density ratios based on column densities reported in 
Table~\ref{tab:cden}. $^a$3$\sigma$ upper limit. Chamaeleon 
and TMC-1 values are from \cite{lis2025}. The bottom three lines show 
revised column density ratios for G300.91+0.88, G304.76+1.34, 
and G309.91+0.32 assuming a higher excitation temperature of 7.0~K, as
described in the text.
\end{table*}

\subsection{Local thermodynamic equilibrium molecular column
  densities}\label{sec:cden}

Since collisional rates are not available for many of the molecules
studied here, we consistently used the LTE approach as implemented in
the Weeds software package. We used molecular spectroscopy data from
the CDMS catalog \citep{muller2001, muller2005} in the calculations.
We assumed the same excitation temperature for all molecules equal to
the ammonia excitation temperature. More details on the Weeds modeling
can be found in \cite{lis2025}. For HC$_3$N, the hyperfine structure
is not included in Weeds. Moreover, while the satellite components are
detected in HOPS-108 (Figure~\ref{fig:hops108}, panel m), only the two
strongest, blended central components are detected in most sources.
Since the emission is optically thin, we performed a Gaussian fit to
the two strongest, blended HFS components and multiplied the column
density by 1.395 to correct for the weak satellite components.

The resulting molecular column densities are listed in
Table~\ref{tab:cden}. These should be interpreted as peak column
densities in a pencil beam, as the convolution of the source size with
the telescope beam is carried out directly in Weeds. The deuterated
species $c$-C$_3$HD is not 
detected in any of the sources studied here. The only molecules detected 
in HOPS--373 are ammonia and C$_3$H$_2$. Sulfur bearing species CCS 
and C$_3$S are not detected in  HOPS--108.

Table~\ref{tab:crat} gives selected column density ratios based on
column densities reported in Table~\ref{tab:cden}. Values for the two
Chamaeleon sources and TMC-1 are from \cite{lis2025}. The
$c$-C$_3$H$_2$/HC$_5$N, HC$_3$N/HC$_5$N, and CCS/C$_3$S ratios are in
general comparable to the values derived in the Chamaeleon cores.
However the $c$-C$_3$HD/$c$-C$_3$H$_2$ upper limits are a factor of 2--3 lower
than in Chamaeleon.

HC$_5$N has three rotational transitions within the DSS-43 frequency
range, with similar upper level energies and excitation requirements.
A comparison of the observed line intensities with the best fit Weeds
models for the individual lines show that the uncertainties in the
molecular column densities, including calibration and modeling, are in
the range 3\% (HOPS-108) up to \about 50\% (BHR71-IRS1, G300.91+0.88,
G304.76+1.34). We estimate the overall uncertainties of the column
density ratios reported in Table~\ref{tab:crat} to be a factor of
2--3.

\subsection{High ammonia column
  densities in G300.91+0.88, G304.76+1.34, and G309.91+0.32}\label{sec:cammo} 

One striking result of the calculations presented above are the very
high ammonia column densities in G300.91+0.88, G304.76+1.34, and
G309.91+0.32, about an order of magnitude higher than those in
Chamaeleon or other sources in our sample. These targets were selected
from the deuterated ammonia survey of \cite{wienen2021}, therefore
they are naturally strong ammonia emitters. However, the high column
densities derived from the Weeds LTE fits require further
investigation. The kinetic temperatures of these three sources,
14--18~K, are in the same range as those derived for other sources,
13--20~K. However, their ammonia excitation temperatures are very low,
4.1--5.2~K, compared to 6--10~K in other sources. This is
understandable, as these sources are more distant
(Table~\ref{tab:param}), thus the DSS-43 beam probes primarily their
lower density, extended envelopes. The resulting $T_{ex}/T_k \sim 0.3$
is low, compared to 0.4--0.76 in other sources. The high column
densities are a direct consequence of the low excitation temperatures,
as the column density is a strong function of the excitation
temperature in the low temperature regime (see red curves in panels d
in Figures~\ref{fig:g300}--\ref{fig:g309}).

The source sizes used in the Weeds fits were derived from the
\textit{Herschel}/SPIRE 350~~$\mu$m dust continuum images. In
particular, G304.76+1.34 has the largest source size in our sample,
$109^{\prime\prime}$, much larger than the DSS-43 beam. To investigate
the dependence of the NH$_3$ column on the assumed source size, we
carried out additional Weeds fits for this source assuming smaller
source sizes of $60^{\prime\prime}$ (a typical value for other
sources) and $41^{\prime\prime}$ (the low Vela~C~IRS 31 41 value). The
best fit for the source size of $60^{\prime\prime}$ results in an
excitation temperature of 5.0~K and an NH$_3$ column density of
$8.1 \times 10^{15}$~cm$^{-2}$, a factor of 1.7 times lower than the
value reported in Table~\ref{tab:cden}. The best fit for the source
size of $41^{\prime\prime}$ results in an excitation temperature of
5.5~K and the same NH$_3$ column density
$8.1 \times 10^{15}$~cm$^{-2}$.

For G300.91+0.32, decreasing the source sizes to $41^{\prime\prime}$
increases the excitation temperature to 5.3~K and decreases the NH$_3$
column density to $1.25 \times 10^{16}$~cm$^{-2}$, a factor of 1.4
lower than the value reported in Table~\ref{tab:cden}. 
For G309.91+0.32, decreasing the
source sizes to $41^{\prime\prime}$ increases the excitation
temperature to 6.5~K and decreases the NH$_3$ column density to
$6.7 \times 10^{15}$~cm$^{-2}$, a factor of 1.5 lower than the value
reported in Table~\ref{tab:cden}. We thus conclude that the assumed
source size has only a moderate effect on the resulting NH$_3$ column
densities and cannot explain the difference between G300.91+0.88,
G304.76+1.34, and G309.91+0.32 and the other sources in our sample.

Given the difference in the opacity of the main NH$_3$ (1,1) hyperfine
component and the satellite components, the lines may not trace the
same gas along the line of sight. The optically thick central
component traces preferentially the surface of the cloud, while
the weaker optically thin satellite components trace all gas along the
line of sight. Our fitting approach assumes a single excitation
temperature. In reality, density and temperature gradients are likely
to exist, resulting in variations of the excitation temperature along
the line of sight, depending on the presence or absence of a central
heating source. We note that high-velocity emission is detected in the
NH$_3$ (3,3) spectra of G300.91+0.88 and G309.91+0.32, suggesting the
presence of embedded sources, which may heat up the central parts of
these cores. However, the NH$_3$ (3,3) line in G304.76+1.34 only shows
a narrow component, suggesting absence of a central source. These
three sources thus likely have different temperature profiles. Yet, in
spite of these differences, they are all characterized by low
excitation temperatures and high NH$_3$ column densities.

To further quantify the sensitivity of the NH$_3$ column density to 
the assumed excitation temperature, we fitted the NH$_3$ (1,1) spectra 
in G300.91+0.88, G304.76+1.34, and G309.91+0.32 assuming an excitation 
temperature of 7~K, an average value for the remaining sources, 
including the two Chamaeleon cores. The resulting NH$_3$ column 
densities are reduced by large factors of 8.2, 11, and 3.4, respectively.
The revised column densities for all molecules are reported in the
bottom three lines of Table~\ref{tab:cden}. The corresponding revised 
abundance ratios are listed in the bottom three lines of 
Table~\ref{tab:crat} and are used for comparison with 
the chemical models.

Detailed radiative transfer models would be required
to better constrain the physical conditions and molecular column 
densities. This would also require additional higher spatial 
resolution mapping observations and is beyond the scope of the 
present paper.

\section{Chemical modeling}
\label{sec:chemmod}

The state-of-the-art gas--grain astrochemical model
\textsc{Dnautilus}, originally introduced by \cite{Majumdar2017}, was
employed to compute the chemical composition of the observed dense
cores using observationally constrained densities and temperatures.
Since its initial implementation, the model has undergone substantial
development, with successive updates to the chemical network presented
in a series of next-generation \textsc{Dnautilus} studies spanning
multiple evolutionary stages of star-forming environments. These
include starless cores \citep{Tasa-Chaveli25}, low-mass star-forming
regions \citep{Majumdar17b}, high-mass star-forming regions
\citep{Li25}, and protoplanetary disks \citep{Kashyap24}, among
others.

\textsc{Dnautilus} is specifically designed to model deuterium
fractionation, including multiple deuterated isotopologs, within both
two-phase (gas and grain surface) and three-phase (gas, grain surface,
and grain bulk mantle) frameworks. The current network includes 1606
gas-phase species, 834 grain-surface species, and 737 grain-mantle
species, connected through 83{,}715 gas-phase reactions, 10{,}967
surface reactions, and 9{,}431 mantle reactions. Tables
\ref{tab:reacts} -- \ref{tab:reacts2} list the dominant formation and
destruction process for the molecules studied here for the best fit
model. The model solves the time-dependent chemical evolution of
molecular abundances in both two-phase mode, where the grain is
treated as chemically homogeneous, and three-phase mode, where the
grain surface and bulk mantle are treated as chemically distinct
reservoirs.

The chemical network incorporates 15 elements, with initial elemental
abundances listed in Table~1 of \cite{Majumdar2017}. Molecular
hydrogen is initially present as H$_2$ and HD, while all other
elements are assumed to be in atomic form. Elements with ionization
potentials below 13.6~eV (C, S, Si, Fe, Na, Mg, Cl, and P) are
initially assumed to be singly ionized. The elemental abundance of
carbon is kept fixed across all models, while four C/O ratios (0.5,
0.7, 0.9, and 1.4) are explored by varying the elemental oxygen
abundance.

All simulations were performed using the three-phase version of
\textsc{Dnautilus}. The cosmic-ray ionization rate was fixed at
$1.3 \times 10^{-17}$ s$^{-1}$. The ranges of densities and
temperatures adopted for each core are listed in
Table~\ref{tab:params} and are based on the observationally derived
values reported in Table~\ref{tab:cden}. In most models, the gas and
dust temperatures are assumed to be equal. The only exceptions are the
Chamaeleon cores, which are discussed separately below.

For Cha C2 and Cha MMS1 the densities derived from the NH$_3$ emission
in this paper are 3.5 $\times$ 10$^3$ -- 1.1 $\times$ 10$^4$ cm$^{-3}$
and 1.7 $\times$ 10$^4$ -- 5 $\times$ 10$^4$ cm$^{-3}$ respectively.
These are lower than reported in \cite{lis2025} where the 350 $\mu$m
emission from \emph{Herschel}/SPIRE in a smaller beam was used to
determine densities of 2 -- 8 $\times$ 10$^{5}$ cm$^{-3}$ for Cha C2
and 3 -- 14 $\times$ 10$^5$ cm$^{-3}$ for Cha MMS1. The H$_2$
densities depend on the tracer (e.g., molecular tracers with different
critical densities trace different regions of the core), and
parameters used to convert the dust optical depth into the column
density. Some of the parameters, e.g., the grain emissivity
coefficient are very uncertain, leading to uncertainties in the number
density. We have therefore considered two sets of models, one using
the ammonia densities derived here, which are representative of the
cloud envelope, and another one using the higher densities (and
temperatures) from \cite{lis2025}, characteristic of the denser gas at
the center. These are denoted "low n" and "high n", respectively, in
Table~\ref{tab:params}. Both sets of models assume a grain temperature
$T_{d}$ of 13.2 K for MMS1 and 13 K for
C2.

The best fit model was determined by calculating the distance of
disagreement \citep{Wakelam24, Wakelam2010}:

\begin{equation}
    D(t) = \frac{1}{N_i} \Sigma | log(X_{mod,i}(t)) - log(X_{obs,i}) |
    \label{eq:D}
\end{equation}
where $N_i$ is the number of observations, $X_{mod,i}(t)$ is the
modeled abundance as a function of time, and $X_{obs,i}$ is the
observed abundance. The best fit model has the lowest $D(t)$ and is
listed in Table~\ref{tab:models}. We take the ratios presented in
Table~\ref{tab:crat} for $X_{obs,i}$ and compare them to the modeled
ratios. The parameters of the best fit models, as determined by the
$D$ parameter, are listed in Table~\ref{tab:models}.

Most cores have similar properties with densities from 2 $\times$
10$^3$ to 2 $\times$ 10$^4$ cm$^{-3}$ and C/O ratios of 0.5 or 0.7.
The exception is TMC-1 which has a very high C/O ratio of 1.4. The
chemical ages determined for each core are also similar and fall
between 0.6 and 5 Myr. The oldest chemical age was found for
G304.76+1.34, while the youngest were BHR71-IRS1, G309.91+0.32 and
Cha2 MMS1. The less dense cores tend to be the ones with the oldest
chemical age, as might be expected given that chemical timescales
scale with density.

\begin{table}[]
\begin{center}
    \caption{\label{tab:params}Parameters used for chemical models of each core.}
    \begin{tabular}{lcccc}
\hline \hline 
  \rule[-3mm]{0mm}{8mm}
    Source & $n$ & $T_{gas}$ & $T_{grain}$\\
  \hline %\rule{0pt}{2ex}
    HOPS-108           & $2\times 10^3 - 1\times 10^4$ & 19.6 & 19.6 \\
    Vela C IRS 31 14   & $2\times 10^3 - 1\times 10^4$ & 15.1 & 15.1 \\
    BHR71-IRS1         & $1\times 10^4 - 5\times 10^4$ & 13.1 & 13.1 \\
    G300.91+0.88       & $2\times 10^3 - 1\times 10^4$ & 14.7 & 14.7 \\
    G304.76+1.34       & $2\times 10^3 - 1\times 10^4$ & 13.7 & 13.7 \\
    G309.91+0.32       & $2\times 10^3 - 1\times 10^4$ & 17.6 & 17.6 \\
    Cha MMS1  (high n)         & $3\times 10^5 - 1.4\times 10^6$  &  8.5 -- 11 & 13.2    \\
    Cha MMS1 (low n)         & $1\times 10^4 - 5\times 10^4$ & 11.0 & 13.2 \\
    Cha C2   (high n)         & $2\times 10^5 - 8\times 10^5$  &  8.5 -- 11  & 13    \\
    Cha C2   (low n)         & $3\times 10^3 - 1\times 10^4$ & 11.0 & 13.0 \\
    TMC-1              & $3\times 10^4$       & 10.0 & 10.0 \\
    \hline %\rule{0pt}{2ex}
\end{tabular}    
\end{center}
Notes: Entries in the table are the source, density (cm$^{-3}$), gas and grain temperatures (K). C/O ratios for each set of parameters are set to 0.5, 0.7, 0.9 and 1.4.
\end{table}

\begin{table}[]
    \begin{center}
    \caption{\label{tab:models}Best fit model parameters for each core.}
    \begin{tabular}{lcccc}
\hline \hline 
  \rule[-3mm]{0mm}{8mm}
  Source & C/O & $n$ & $T_{gas}$ & $t$ \\ 
  \hline %\rule{0pt}{2ex}
    HOPS-108          & 0.7 & $8 \times 10^3$ & 19.6 & $8.0 \times 10^5$ \\
    Vela C IRS 31 14  & 0.5 & $2 \times 10^3$ & 15.1 & $5.0 \times 10^6$ \\
    BHR71-IRS1        & 0.7 & $1 \times 10^4$ & 13.1 & $6.3 \times 10^5$ \\
    G300.9+0.88       & 0.7 & $2 \times 10^3$ & 14.7 & $2.5 \times 10^6$ \\
    G304.76+1.34      & 0.5 & $2 \times 10^3$ & 13.7 & $5.0 \times 10^6$ \\
    G309.91+0.32      & 0.7 & $1 \times 10^4$ & 17.6 & $6.3 \times 10^5$ \\
    Cha MMS1 (high n)& 0.7 & $3 \times 10^5$ & 10.0 & $6.3 \times 10^4$ \\
    Cha MMS1 (low n) & 0.7 & $2 \times 10^4$ & 11.0 & $6.3 \times 10^5$ \\
    Cha C2  (high n) & 0.5 & $2 \times 10^5$ & 8.5 &  $8.0 \times 10^4$ \\
    Cha C2  (low n)  & 0.5 & $8 \times 10^3$ & 11.0 & $1.0 \times 10^6$ \\
    TMC-1             & 1.4 & $3 \times 10^4$ & 10.0 & $6.3 \times 10^5$ \\
 \hline 
\end{tabular}
\end{center}
Notes: Entries in the table are the source, C/O ratio, density
(cm$^{-3}$), gas temperature (K), and time (yr). The best fit models
are determined by the minimum value of the distance of disagreement
(D) parameter (Eq.~\ref{eq:D}). 
\end{table}

Direct comparisons of the chemical ages and elemental C/O ratios
derived for the majority of the dense cores in this work with previous
studies remain limited, primarily because detailed gas-grain chemical
modeling is available only for a small number of well-studied sources,
most notably TMC-1
\citep[e.g.,][]{Majumdar2017,Navarro21,Tasa-Chaveli25}. Previous
kinetic models of TMC-1 generally infer chemical ages of order
$0.1 - 6$~Myr, depending on the adopted chemical network, elemental
abundances, and physical conditions
\citep[e.g.,][]{Majumdar2017,Navarro21,Wakelam24,Maitrey25}. Our
derived chemical ages of $\sim0.6$--$5$ Myr are broadly consistent
with the range expected for cold dense cores, particularly considering
that chemical timescales scale with density \citep{Agundez13}.

The C/O ratios inferred for most of the sources in our sample
($0.5-0.7$) are also consistent with the elemental abundances commonly
adopted in dense cloud chemical models based on diffuse cloud
depletion studies toward $\zeta$~Oph \citep{Jenkins09,Hincelin2011}.
The higher C/O ratio derived for TMC-1 in our modeling is also
consistent with previous studies of this source, showing that enhanced
C/O ($\gtrsim1$) increases the production of long carbon-chain
molecules and cyanopolyynes toward the cyanopolyyne peak
\citep{loomis21,burkhardt21,byrne26}. The sensitivity of modeled
abundances to the adopted C/O ratio has long been recognized in
dense-core chemical models, particularly for TMC-1, where proposed
elemental C/O ratios span a wide range from $\sim0.55$ to $\sim1.4$
\citep{Agundez13}. However, detailed chemical modeling studies for
other dense cores observed in our survey remain scarce. Therefore, the
chemical ages derived here provide useful additional constraints on
the evolutionary stages of such dense cores. It should also be noted
that inferred chemical ages depend on assumptions regarding the
adopted chemical network, initial elemental abundances, and chemical
desorption processes, which can shift the best-fit times by factors of
a few \citep{Wakelam24,Maitrey25}.

\begin{figure*}[!ht]
    \centering
    \includegraphics[width=0.3\linewidth,trim={70 20 100 50},clip]{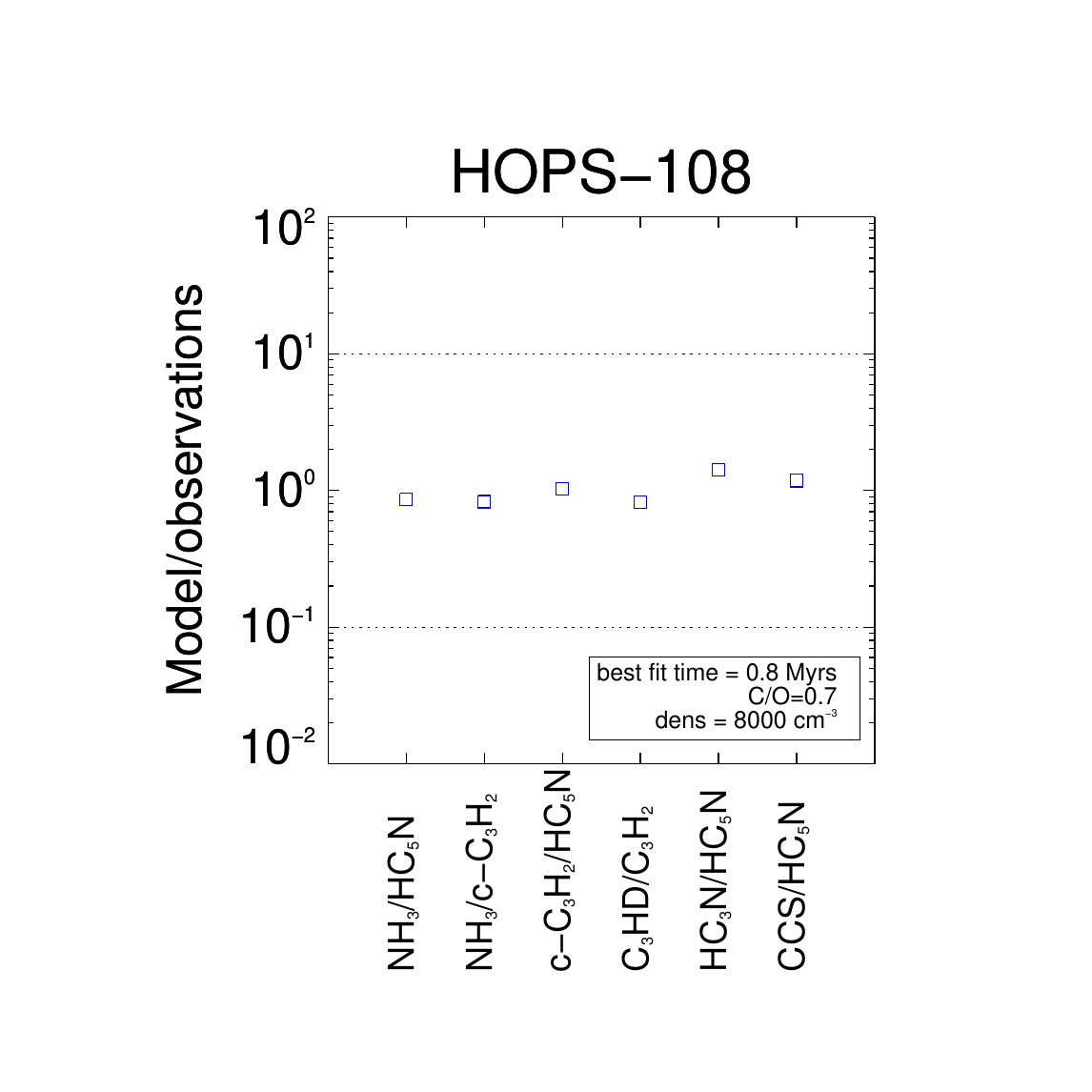}
    \includegraphics[width=0.3\linewidth,trim={70 20 100 50},clip]{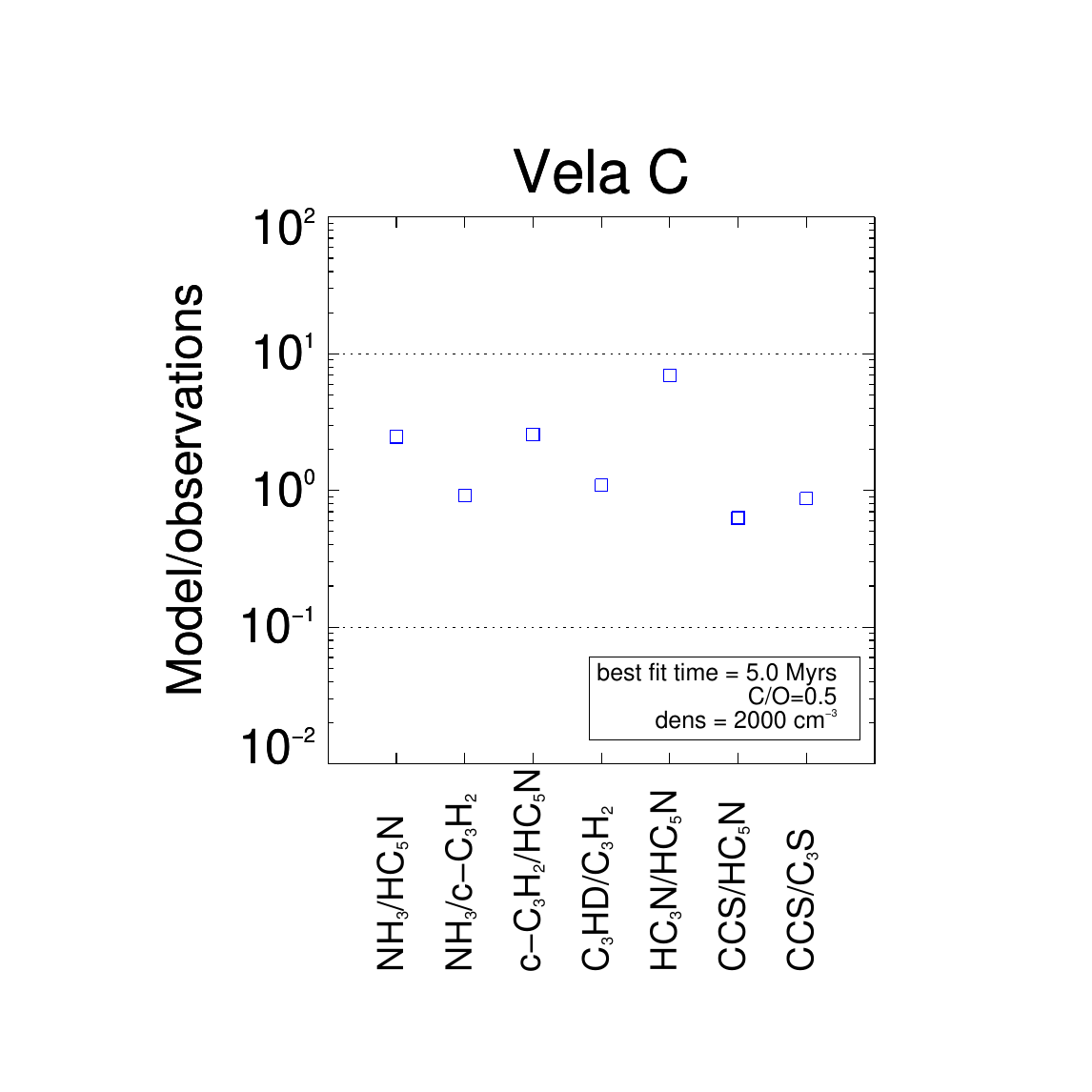}
    \includegraphics[width=0.3\linewidth,trim={70 20 100 50},clip]{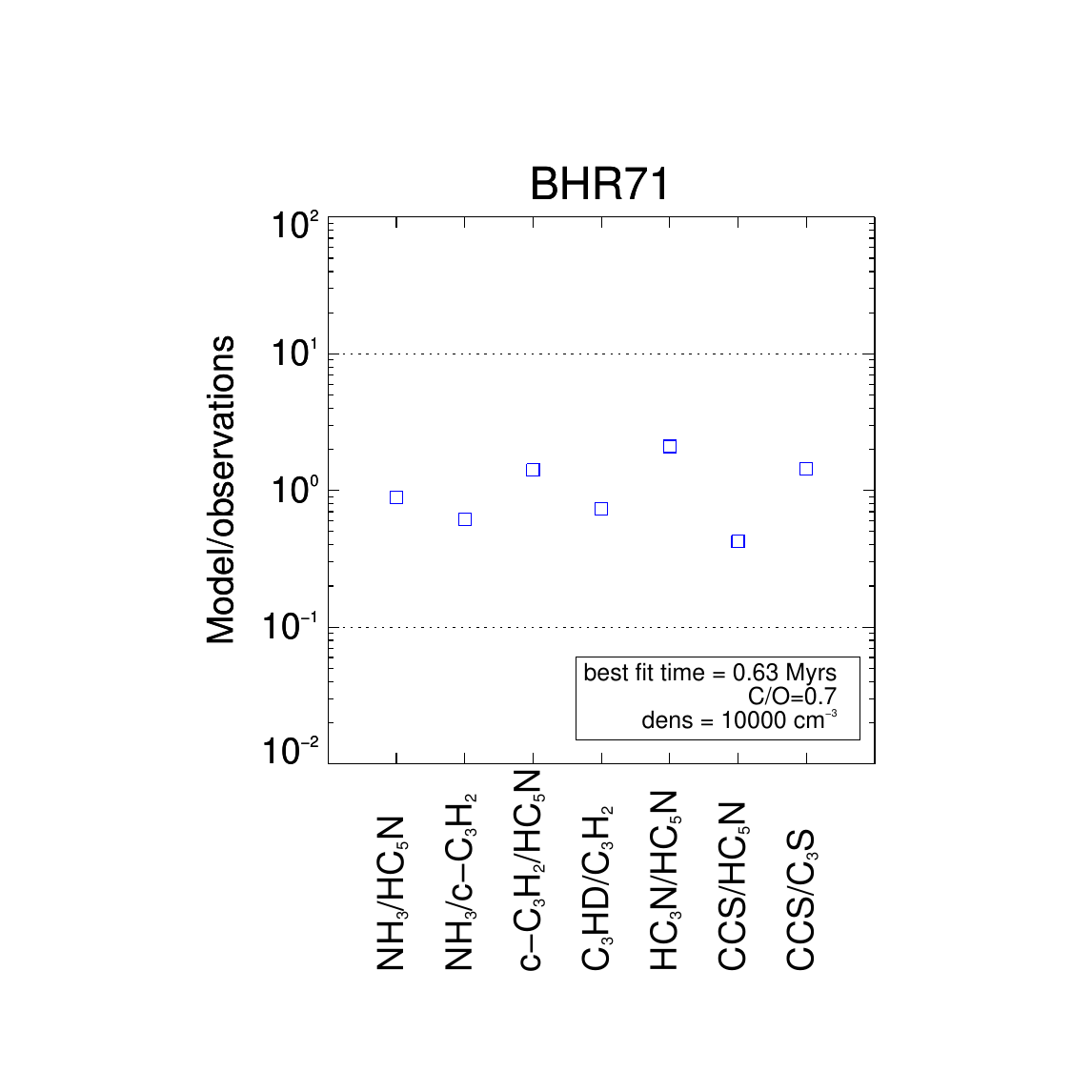}\\
    \vspace{0.2cm}
    \includegraphics[width=0.3\linewidth,trim={70 20 100 50},clip]{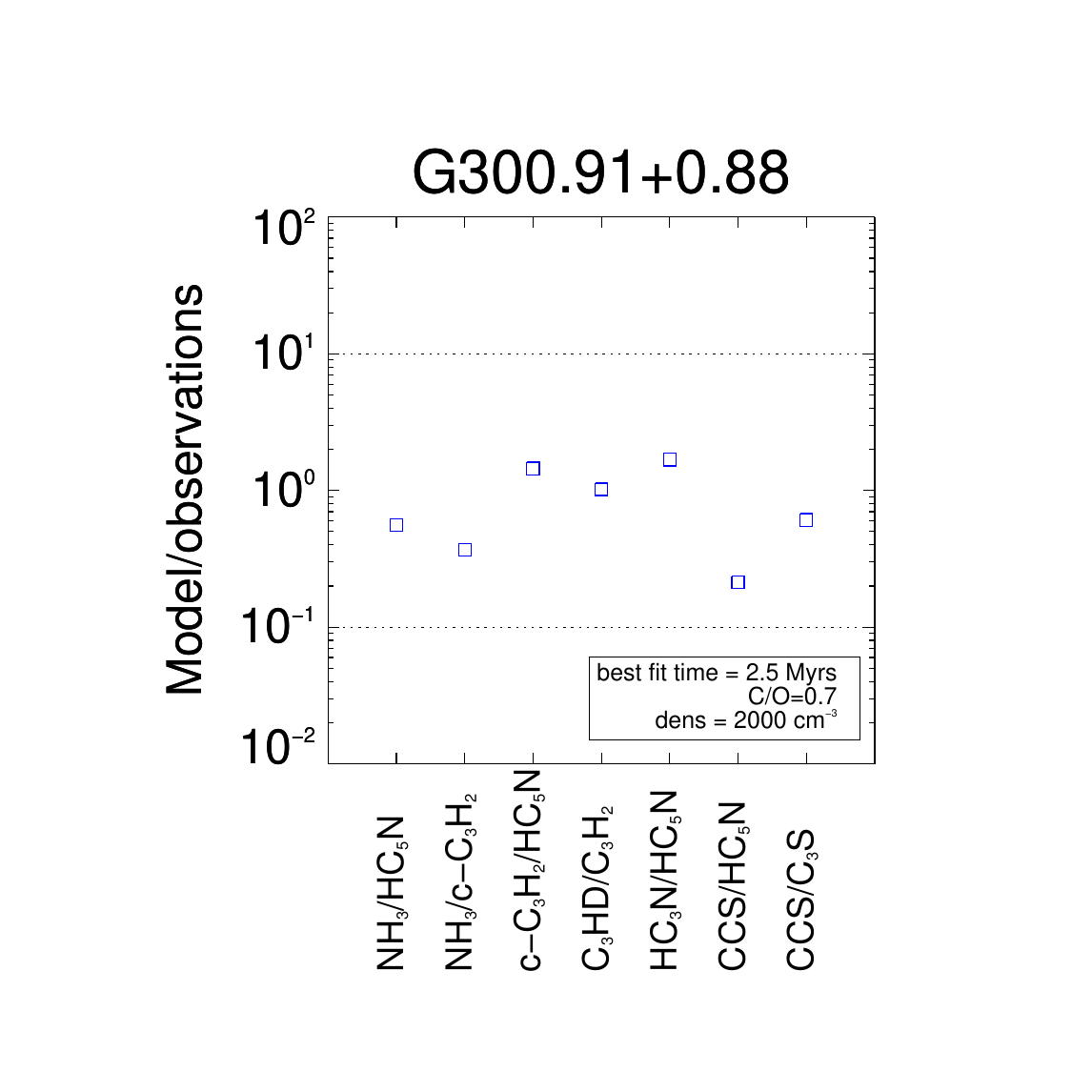}
    \includegraphics[width=0.3\linewidth,trim={70 20 100 50},clip]{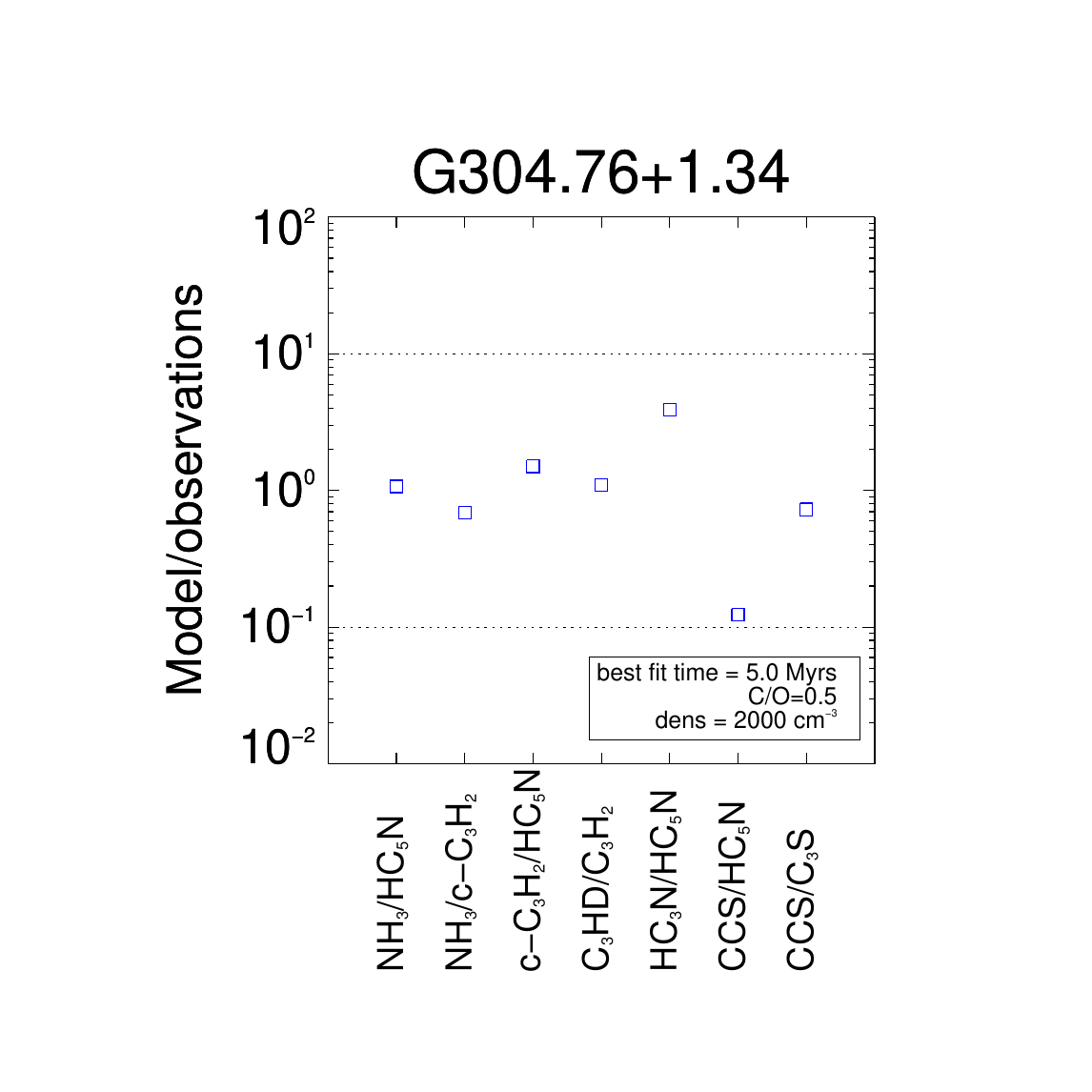}
    \includegraphics[width=0.3\linewidth,trim={70 20 100 50},clip]{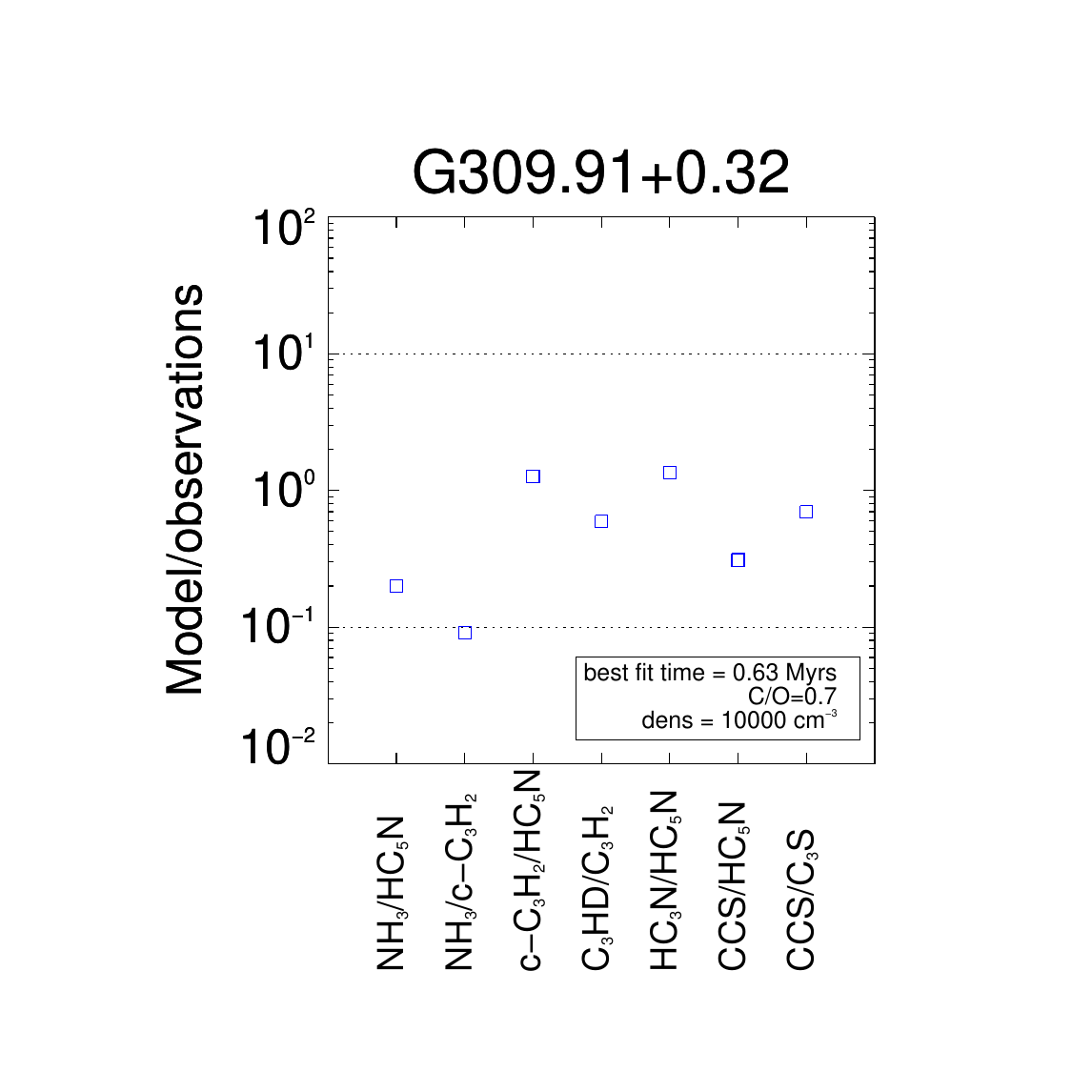}\\
    \vspace{0.2cm}
    \includegraphics[width=0.3\linewidth,trim={70 20 100 50},clip]{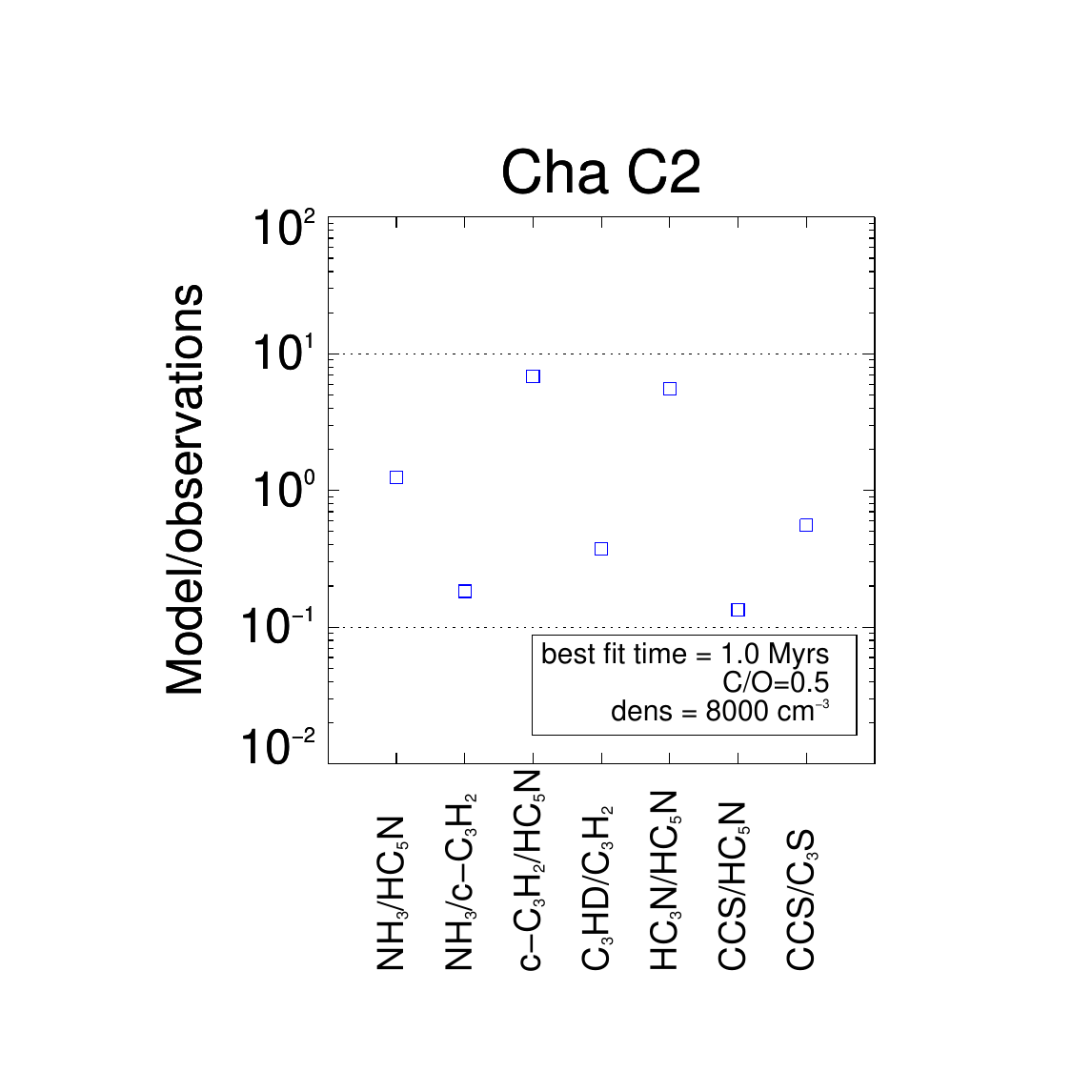}
    \includegraphics[width=0.3\linewidth,trim={70 20 100 50},clip]{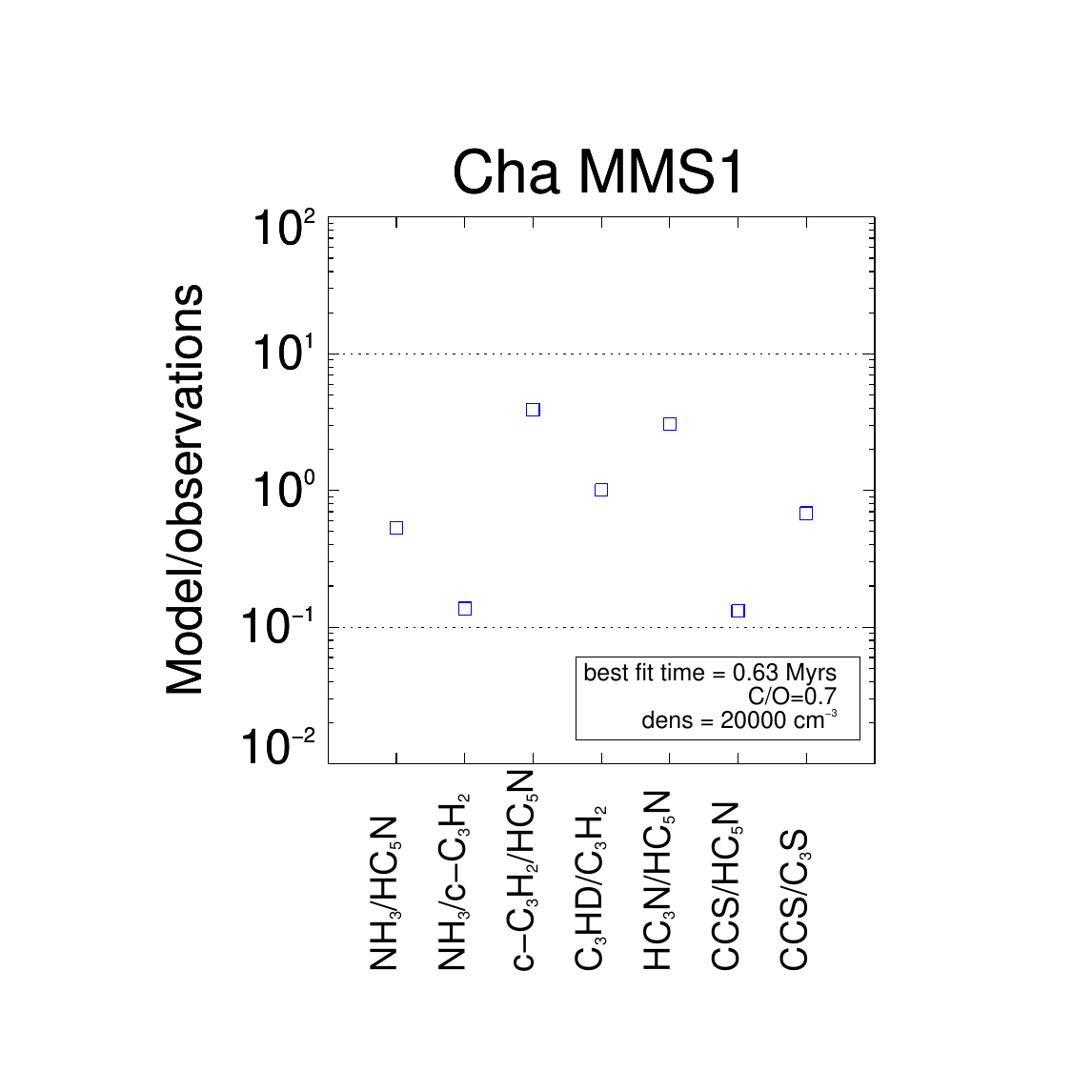}
    \includegraphics[width=0.3\linewidth,trim={70 20 100 50},clip]{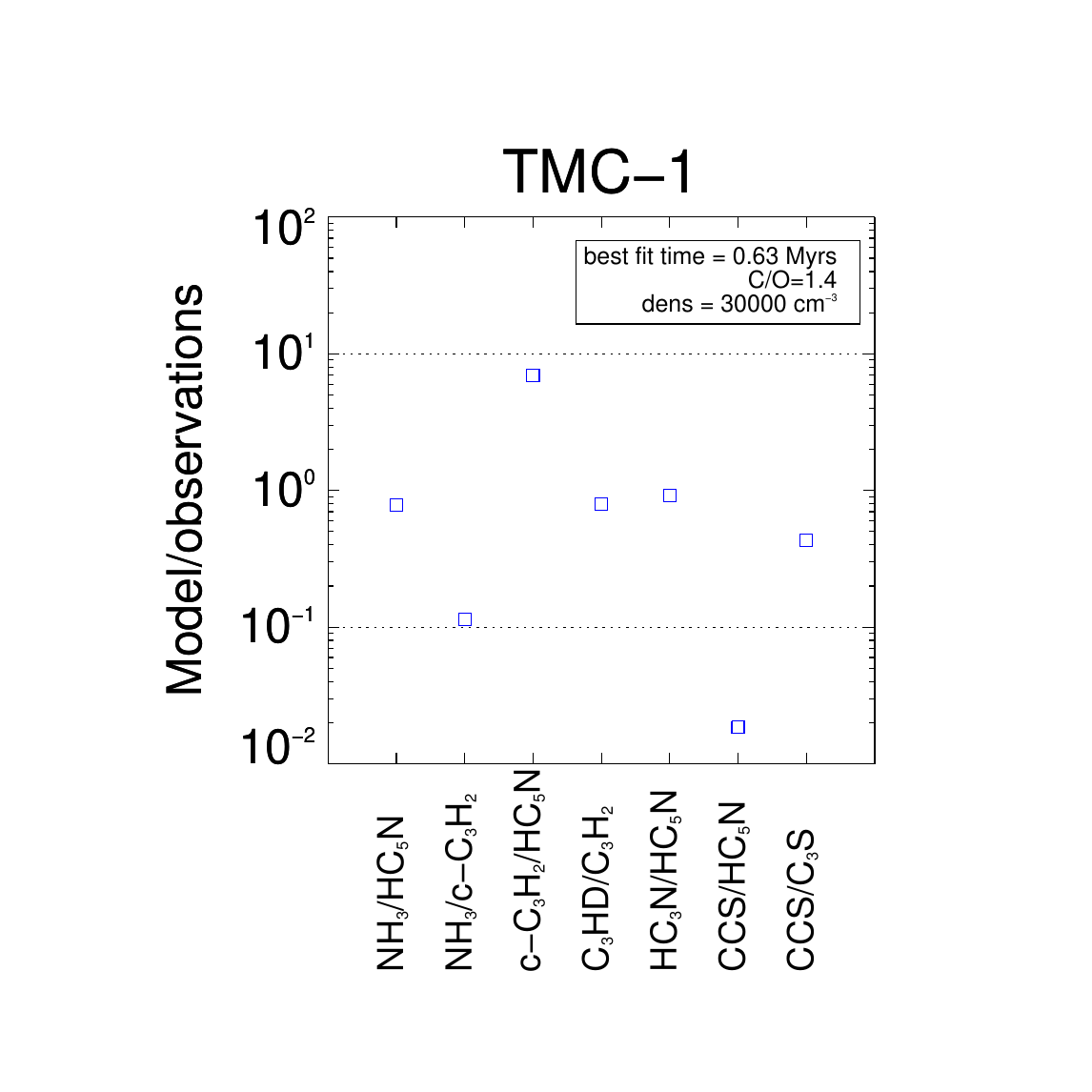}
    \caption{Models ratios/observed ratios for best fit model for each core. The horizontal dashed lines show ratios of 10 and 0.1. Cha C2 and MMS1 models assume the "low n" conditions derived in this paper.}

    \label{fig:ratios}
\end{figure*}

Figure~\ref{fig:ratios} compares the observed molecular abundance
ratios with those predicted by our best-fit models, expressed as the
ratio of observed to modeled values. For most species, the agreement
is within a factor of 10. A notable exception is the CCS/HC$_3$N
ratio, for which the model underpredicts the observed ratio. This
discrepancy likely reflects a combination of the longstanding ``sulfur
depletion'' problem in astrochemistry \citep{Majumdar16, Vidal17} and
limitations in reproducing early-time carbon-chain chemistry. 
Adopting an elemental sulfur abundance significantly below its cosmic
value (a few $\times 10^{-5}$; \citealt{Jenkins09}) is commonly
required in dense-cloud models to avoid overproducing the observed
S-bearing species \citep{Majumdar16, Vidal17}. In this work, we adopt
a depleted sulfur abundance of $8\times10^{-8}$, a value commonly used
in dense core chemical models \citep{Majumdar16, Majumdar2017,
  Navarro21, Wakelam24, Maitrey25}. This may suppress the formation
efficiency of sulfur-bearing carbon chains such as CCS. \citet{seo19}
demonstrated the strong sensitivity of the CCS abundance on the
assumed initial sulfur abundance in dense cores of the Taurus
molecular cloud. In addition, different degrees of sulfur depletion
may be present across dense cores \citep{seo19, Vastel18}.

In practice, a too-low modeled CCS/HC$_3$N ratio indicates either an
underproduction of CCS or an overproduction of HC$_3$N under the
adopted physical and chemical conditions. Since CCS is directly
sensitive to the available gas-phase sulfur reservoir, whereas HC$_3$N
primarily traces carbon- and nitrogen-chain chemistry, enhanced sulfur
depletion preferentially reduces CCS, thereby lowering the CCS/HC$_3$N
ratio. In addition, the balance between carbon-chain growth and
oxidation pathways, regulated by the elemental C/O ratio, plays a key
role in controlling the relative abundances of CCS and HC$_3$N.

Moreover, the use of a single-point physical model may not adequately
capture the spatial differentiation between CCS- and HC$_3$N-emitting
regions. CCS is expected to preferentially trace relatively diffuse
and chemically younger layers, while HC$_3$N is more likely to arise
from denser and better-shielded gas. Such chemical and physical
stratification can further contribute to the observed discrepancy.

Future work incorporating more realistic physical structures and
systematically exploring a range of initial sulfur abundances,
spanning different degrees of depletion, will be essential to
reconcile the modeled and observed CCS/HC$_3$N ratios.

\section{Summary}
\label{sec:conclusion}

With this work, we extended our 1.3~cm wavelength radio survey for
organics in the southern hemisphere by observing seven dense cores
selected from the ALMA COMPASS program (HOPS-108, HOPS-373, BHR-71)
and the deuterated ammonia survey of \cite{wienen2021} (G300.91+0.88,
G304.76+1.34, G309.91+0.32) using the 70-m DSS-43 antenna of NASA's
Deep Space Network. Prior observations of two Chamaeleon cores, MMS1
and C2 \citep{lis2025}, are also included in the analysis.

The main results can be summarized as follows:
\begin{itemize}
  \item Observations of the NH$_3$ (1,1) inversion line were used to
    determine the excitation temperatures, as well as the average gas
    kinetic temperatures, which vary between $\sim 11$ and 20~K.   
  \item LVG models of the NH$_3$ (1,1) inversion line were then used
    to determine the average H$_2$ densities. The results are
    sensitive to the collisional cross sections used, and vary among
    sources from $(1-5) \times 10^4$ cm$^{-3}$ in BHR-71 to $(2 - 10)
  \times 10^3$~cm$^{-3}$ in the more distant sources from the
    deuterated ammonia sample. 
  \item Molecular column densities (and upper limits) of NH$_3$,
    $c$-C$_3$H$_2$, HC$_3$N, HC$_5$N, CCS, C$_3$S, and $c$-C$_3$HD were
    derived for each source under the assumption of LTE. The resulting
    column density ratios were compared with predictions of the
    state-of-the-art gas--grain astrochemical model \textsc{Dnautilus}
    \citep{Majumdar2017}. These comparisons were used to constrain the
    elemental C/O ratio and the chemical age of each source.
  \item Most cores show similar C/O ratios in the range $0.5-0.7$,
    significantly lower than the well-studied carbon-chain-rich dense
    core TMC-1, which is characterized by C/O $\sim 1.4$ and enhanced
    hydrocarbon chemistry.
  \item The inferred chemical ages span $\sim 0.6 -5$~Myr across the
    sample. The oldest chemical age is found for G304.76+1.34, while
    BHR71-IRS1, G309.91+0.32, and Cha2 MMS1 are chemically younger.

\end{itemize}

Our results showcase the synergistic approach of combining
radio observations using the DSS-43 antenna with state-of-the-art
astrochemical models to study the chemical composition of southern
hemisphere dense cores, as a complement to GBT observations of
northern sources.

\begin{acknowledgements}
  This research was carried out at the Jet Propulsion Laboratory,
  California Institute of Technology, under a contract with the
  National Aeronautics and Space Administration (80NM0018D0004) and
  funded through the internal Research and Technology Development
  program. We thank Steve Lichten, Joe Lazio, and the DSN staff for
  their support and assistance with the DSS-43 observations. LM
  acknowledges funding support from the DAE through the NISER project
  RNI 4011.
\end{acknowledgements}

\begin{appendix}

\section{DSS-43 spectra of the remaining sources and dominant formation and
  destruction processes}\label{app:spectra} 

\begin{figure}[ht!]
\centering
\includegraphics[trim=3.2cm 2.5cm 2.3cm 5cm, clip=true, %angle=90,
width=0.89\columnwidth,angle=0]{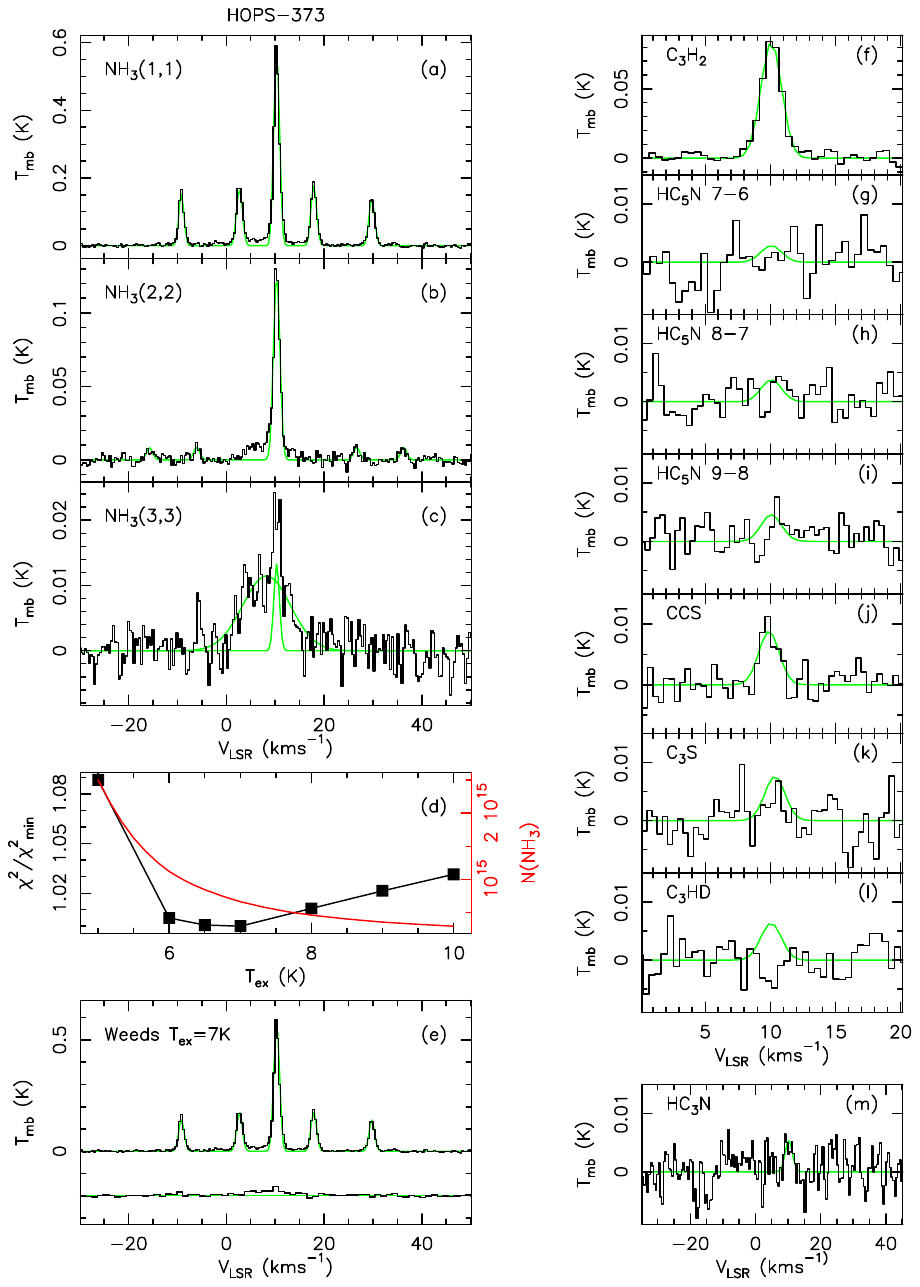} %L B R T 
\caption{DSS-43 observations of HOPS-373.} 
\label{fig:hops373}
\end{figure}

\begin{figure}[ht!]
\centering
\includegraphics[trim=3.2cm 2.5cm 2.3cm 5cm, clip=true,  %angle=90,
width=0.89\columnwidth,angle=0]{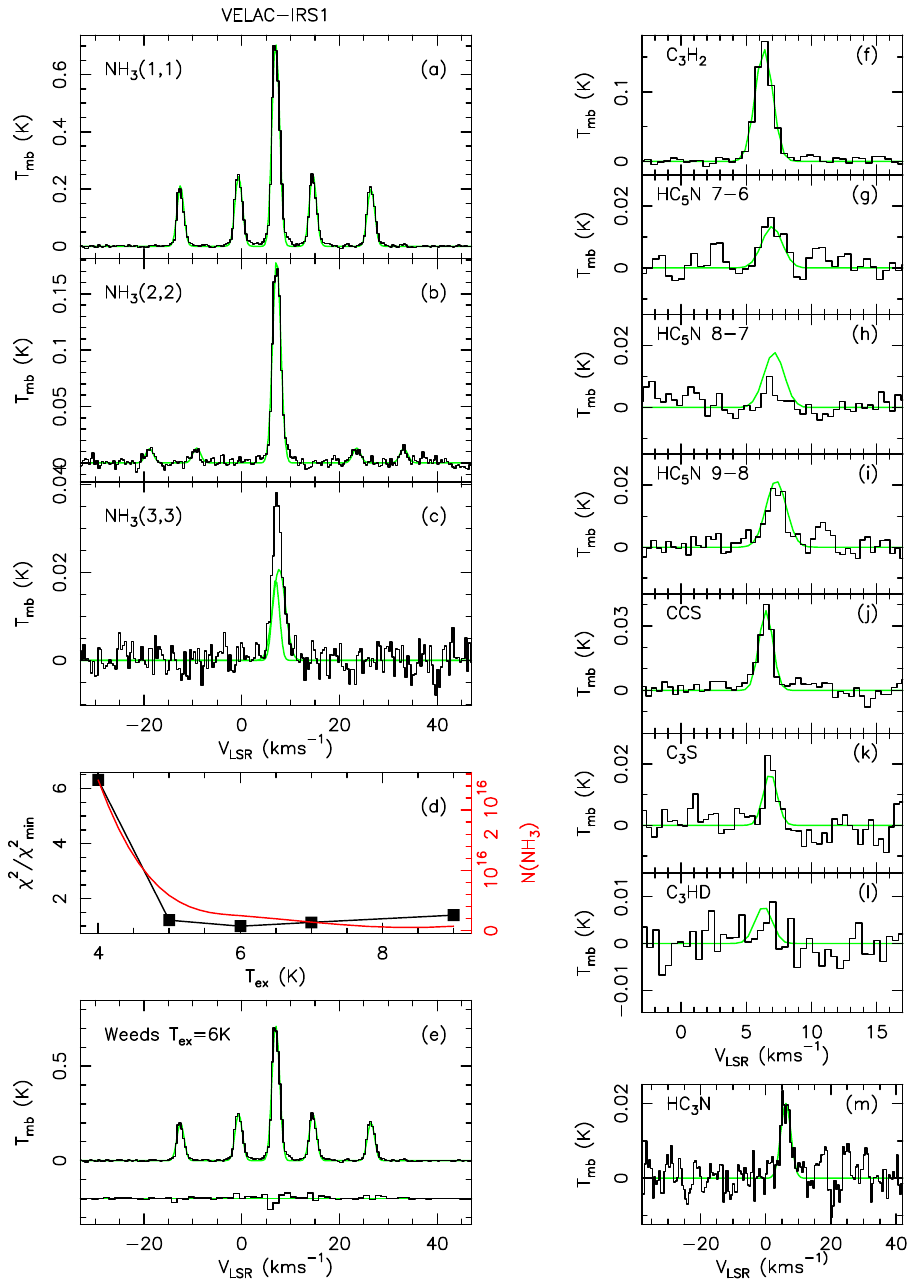} %L B R T 
\caption{DSS-43 observations of Vela C IRS1.} 
\label{fig:velac}
\end{figure}

\begin{figure}[ht!]
\centering
\includegraphics[trim=3.2cm 2.5cm 2.3cm 5cm, clip=true,  %angle=90,
width=0.89\columnwidth,angle=0]{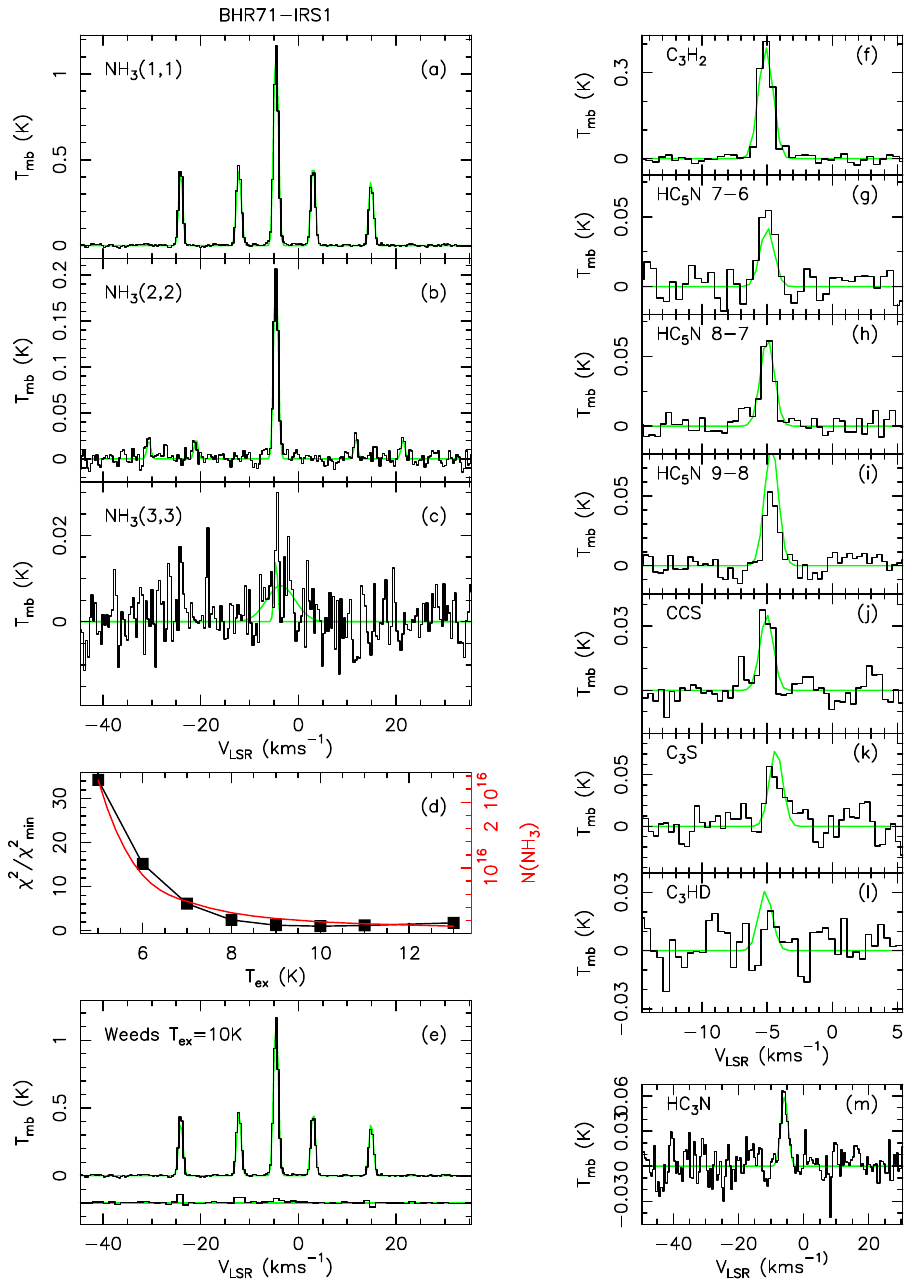} %L B R T 
\caption{DSS-43 observations of BHR71-IRS1.} 
\label{fig:bhr71}
\end{figure}

\begin{figure}[ht!]
\centering
\includegraphics[trim=3.2cm 2.5cm 2.3cm 5cm, clip=true,  %angle=90,
width=0.89\columnwidth,angle=0]{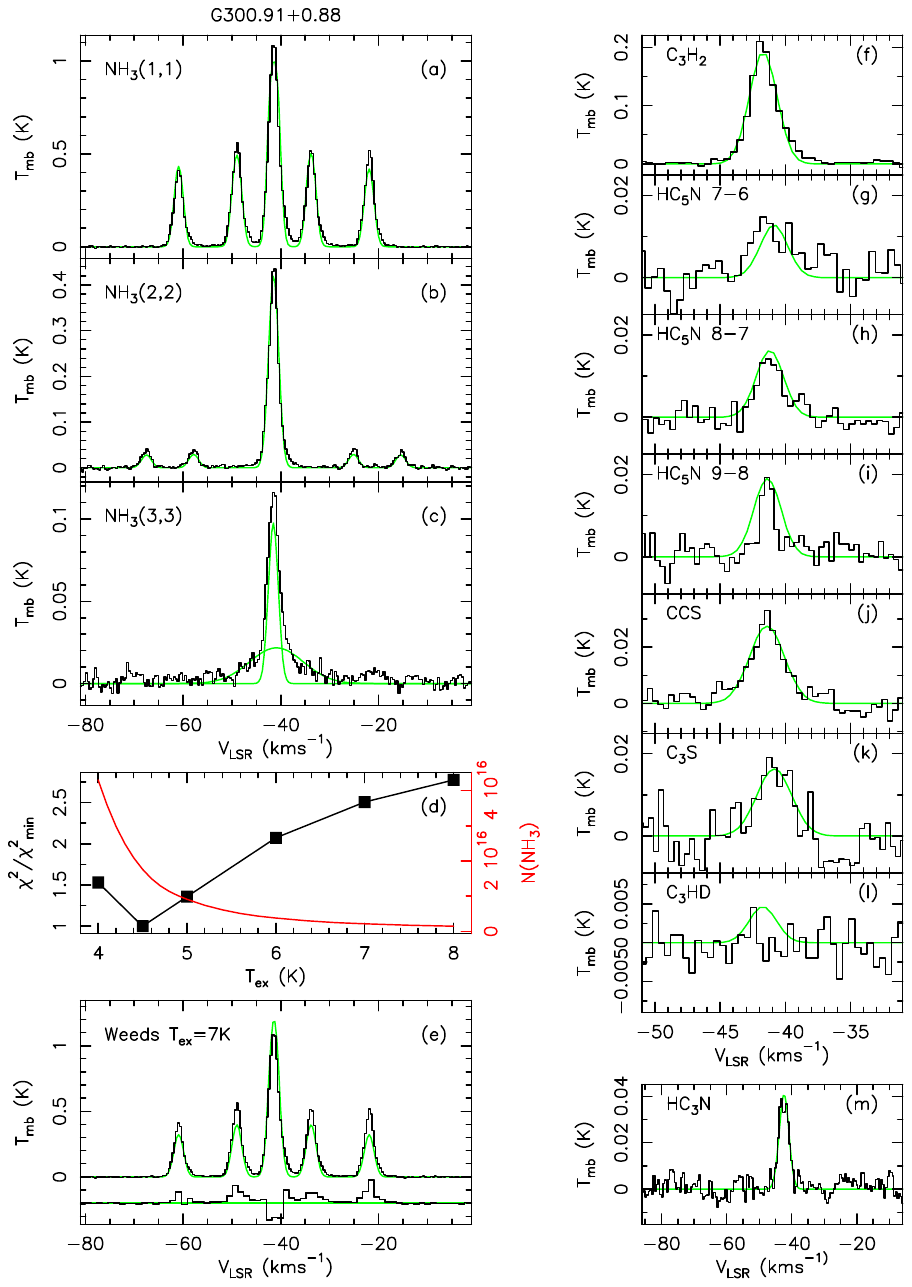} %L B R T 
\caption{DSS-43 observations of G300.91+0.88.} 
\label{fig:g300}
\end{figure}

\begin{figure}[ht!]
\centering
\includegraphics[trim=3.2cm 2.5cm 2.3cm 5cm, clip=true,  %angle=90,
width=0.89\columnwidth,angle=0]{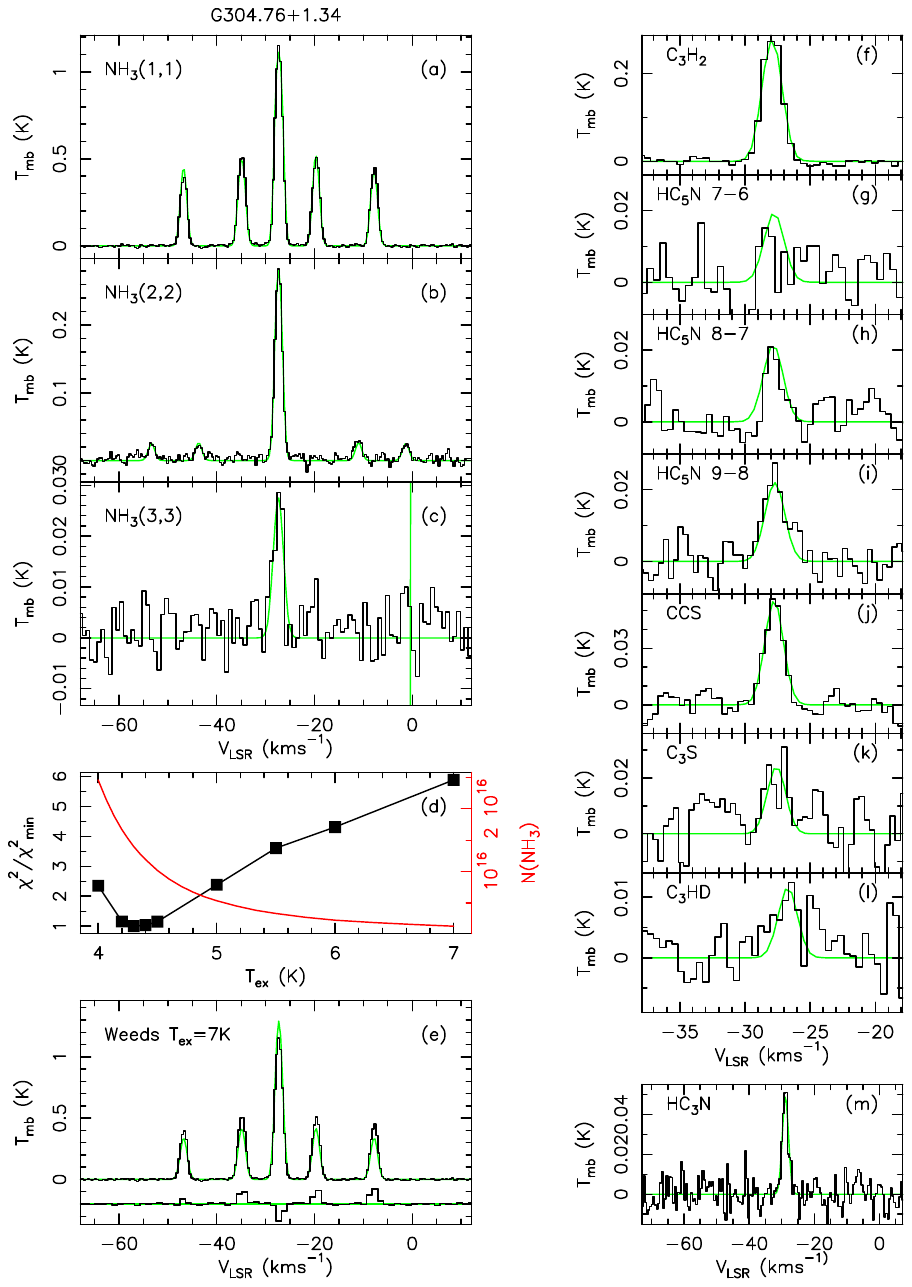} %L B R T 
\caption{DSS-43 observations of G304.76+1.34.} 
\label{fig:g304}
\end{figure}

\begin{figure}[ht!]
\centering
\includegraphics[trim=3.2cm 2.5cm 2.3cm 5cm, clip=true,  %angle=90,
width=0.89\columnwidth,angle=0]{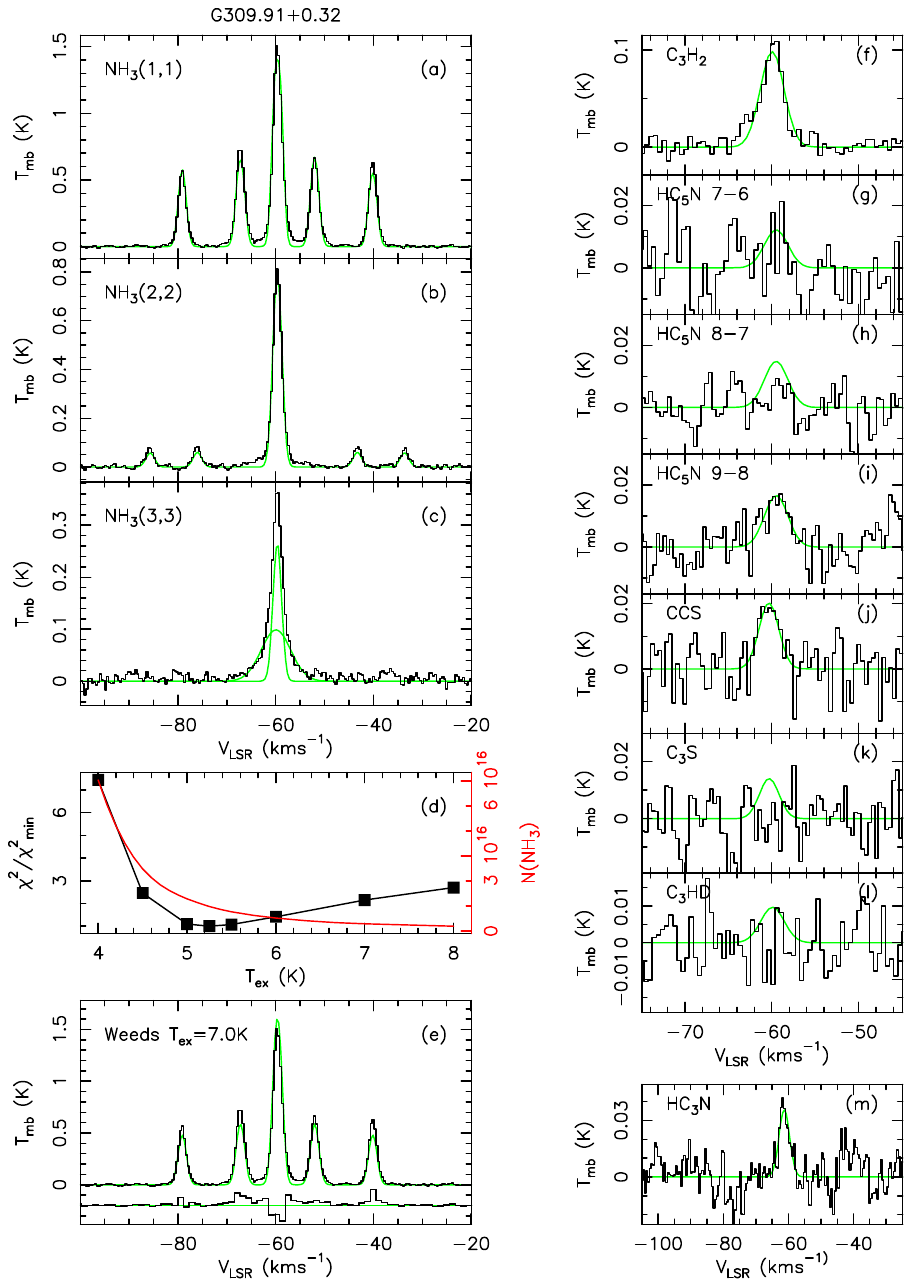} %L B R T 
\caption{DSS-43 observations of G309.91--0.25.} 
\label{fig:g309}
\end{figure}

Figures~\ref{fig:hops373}--\ref{fig:g309} show spectra of the remaining sources
observed with DSS-43. The format is the same as in
Figure~\ref{fig:hops108}. The deuterated species $c$-C$_3$HD is not detected in any of the
sources. In HOPS-373, only ammonia, C$_3$H$_2$, and CCS are detected.
In Vela C, all lines other than $c$-C$_3$HD are detected. In BHR71-IRS1,
C$_3$S is located in a noisy part of the spectrum. Although the
feature is formally 5.4$\sigma$, owing to the difference in the
central velocity and width compared to CCS, we consider it an upper
limit. Weeds fits in Figures~\ref{fig:g300}--\ref{fig:g309} correspond to 
an excitation temperature of 7~K rather than the best fit values. 

Tables \ref{tab:reacts} -- \ref{tab:reacts2} list the dominant
formation and destruction process for each molecule in each core.

\begin{landscape}
    \begin{table}[]
    \tiny
        \centering
          \caption{Dominant formation and destruction processes for
            each molecule in each core.  Reactions are included if
            they contribute more than 10\% to the formation or
            destruction.} 
        \label{tab:reacts}
        \begin{tabular}{lccccccccccc}
             Reaction & HOPS108 & Vela C & BHR71 & G300.9 & G304.76 & G309.91 & \multicolumn{2}{c}{Cha2 MMS1} & \multicolumn{2}{c}{Cha2 C2} & TMC-1 \\
             & & & & & & & high n & low n & high n & low n & \\
\hline
\rule{-2pt}{3ex} 
{\bf NH$_3$ (production)}\\
NH$_4^+$ + e $\rightarrow$ NH$_3$ + H                 & $\checkmark$ & $\checkmark$ & $\checkmark$ & $\checkmark$ & $\checkmark$ & $\checkmark$ & $\checkmark$ & $\checkmark$ & $\checkmark $ & $\checkmark$ & $\checkmark$ \\
NH$_3$D$^+$ + $\rightarrow$ NH$_3$ + D                &     &      &       &      &     &      & $\checkmark$ &      & $\checkmark$  &    & \\
\\
{\bf NH$_3$ (destruction)} \\
NH$_3$ + H$_3^+$ $\rightarrow$ NH$_4^+$ + H$_2$               & $\checkmark$ & $\checkmark$ & $\checkmark$ & $\checkmark$ & $\checkmark$ & $\checkmark$ &     & $\checkmark$ & $\checkmark$ & $\checkmark$ & $\checkmark$\\
NH$_3$ + C$^+$ $\rightarrow$ HCNH$^+$ + H             &      &      & $\checkmark$ & $\checkmark$ &     & $\checkmark$ &      &     &      &      &             \\ 
NH$_3$ + HCO$^+$ $\rightarrow$ NH$_4^+$ + CO          &       &      & $\checkmark$ &     &     &  $\checkmark$ &      &     &       & $\checkmark$ & $\checkmark$ \\
NH$_3$ + H$^+$ $\rightarrow$ NH$_3^+$ + H                 &        &      & $\checkmark$ & $\checkmark$ &    &  $\checkmark$ & $\checkmark$  & $\checkmark$ &     &      &             \\
NH$_3$ + H$_2$D$^+$ $\rightarrow$ NH$_3$D$^+$ + H$_2$      &      &        &     &      &     &      &       & $\checkmark$ & $\checkmark$ &    &              \\
NH$_3$ + CN $\rightarrow$ HCN + NH$_2$              &      &       &     &        &    &      &       &      &     &      & $\checkmark$ \\
NH$_3$ + D$_3^+$ $\rightarrow$ NH$_2$D$_2^+$ + HD          &       &      &      &       &     &     & $\checkmark$  &      & $\checkmark$ &     &              \\
\\
\hline 
\\
{\bf c-C$_3$H$_2$ (production)}\\
c-C$_3$H$_3^+$ + e $\rightarrow$ c-C$_3$H$_2$ + H               & $\checkmark$ & $\checkmark$ & $\checkmark$ & $\checkmark$ & $\checkmark$ & $\checkmark$ & $\checkmark$ & $\checkmark$ & $\checkmark$ & $\checkmark$ & $\checkmark$\\
CH2CCH + H $\rightarrow$ c-C$_3$H$_2$ +  H$_2$         & $\checkmark$ &      &       &     &      &       & $\checkmark $ &     & $\checkmark$ & $\checkmark$ & \\
C$_5$H$_3^+$ + e $\rightarrow$ c-C$_3$H$_2$ + CCH             &       &     &       &      &     &      &       &     &     &       & $\checkmark$\\
c-C$_3$H$_2$D$^+$ + e $\rightarrow$ c-C$_3$H$_2$ + D          &     &        &       &     &     &      & $\checkmark$  &     & $\checkmark$ &       & \\
\\
{\bf c-C$_3$H$_2$ (destruction)}\\
c-C$_3$H$_2$ + H$_3^+$ $\rightarrow$ c-C$_3$H$_3^+$ + H$_2$               & $\checkmark$ & $\checkmark$ & $\checkmark$ & $\checkmark$ & $\checkmark$ & $\checkmark$ &      & $\checkmark$ & $\checkmark$ & $\checkmark$ & $\checkmark$\\
c-C$_3$H$_2$ + HCO$^+$ $\rightarrow$ CO + c-C$_3$H$_3^+$          &      &      & $\checkmark$ &     &      & $\checkmark$ &      &      &      & $\checkmark$ & $\checkmark$\\
c-C$_3$H$_2$ + C$^+$ $\rightarrow$ C$_4^+$ + H$_2$                        &      &      & $\checkmark$ & $\checkmark$ &     &      &      &      &      &       & \\
c-C$_3$H$_2$ + C$^+$ $\rightarrow$ C$_4$H$^+$ + H                     &      &      & $\checkmark$ & $\checkmark$ &     &      &      &      &      &      & \\
c-C$_3$H$_2$ + H$_2$D$^+$ $\rightarrow$ c-C$_3$H$_2$D$^+$ + H$_2$     &       &      &     &      &     &      &      & $\checkmark$ & $\checkmark$ &      & $\checkmark$\\
c-C$_3$H$_2$ + D$_3^+$ $\rightarrow$ c-C$_3$HD$_2^+$ + HD           &       &      &      &      &     &      & $\checkmark$ &      & $\checkmark$ &      & \\
c-C$_3$H$_2$ + D$_3^+$ $\rightarrow$ c-C$_3$H$_2$D$^+$ + H$_2$          &       &      &      &      &     &      & $\checkmark$ &      &      &      & \\
\hline
\\
{\bf c-C$_3$HD (production)}\\
c-C$_3$H$_2$D$^+$ + e $\rightarrow$ c-C$_3$HD + H      & $\checkmark$ & $\checkmark$ & $\checkmark$ & $\checkmark$ & $\checkmark$ & $\checkmark$ & $\checkmark$ & $\checkmark$ & $\checkmark$ & $\checkmark$ & $\checkmark$ \\
c-C$_3$HD$_2^+$ + e $\rightarrow$ c-C$_3$HD + H        &              &              &              &              &              &              & $\checkmark$ & $\checkmark$ & $\checkmark$ &              & \\
C$_5$H$_2$D$^+$ + e $\rightarrow$ c-C$_3$HD + CCH      &              &              &              &              &              &              &                 &              &              &              & $\checkmark$\\
\\
{\bf c-C$_3$HD (destruction)}\\
c-C$_3$HD + H$_3^+$ $\rightarrow$ c-C$_3$H$_2$D$^+$ + H$_2$ & $\checkmark$ & $\checkmark$ & $\checkmark$ & $\checkmark$ & $\checkmark$ & $\checkmark$ &              & $\checkmark$ &              & $\checkmark$ & $\checkmark$ \\
c-C$_3$HD + H$_3^+$ $\rightarrow$ c-C$_3$H$_3^+$ + HD       & $\checkmark$ & $\checkmark$ & $\checkmark$ & $\checkmark$ & $\checkmark$ & $\checkmark$ &              & $\checkmark$ &              & $\checkmark$ &  $\checkmark$ \\
c-C$_3$HD + HCO$^+$ $\rightarrow$ c-C$_3$HD$^+$ + CO        &              &              & $\checkmark$ &              &              & $\checkmark$ &              &              &              & $\checkmark$ & $\checkmark$\\
c-C$_3$HD + C$^+$ $\rightarrow$ C$_4^+$ + HD                &              &              & $\checkmark$ & $\checkmark$ &              &              &              &              &              &              &              \\
c-C$_3$HD + H$_2$D$^+$ $\rightarrow$ c-C$_3$H$_2$D$^+$ + HD &              &              &              &              &              &              &              & $\checkmark$ & $\checkmark$ &              & $\checkmark$\\
c-C$_3$HD + D$_3^+$ $\rightarrow$ c-C$_3$H$_2$D$^+$ + D$_2$ &              &              &              &              &              &              & $\checkmark$ &              & $\checkmark$ &              &              \\
c-C$_3$HD + D$_3^+$ $\rightarrow$ c-C$_3$HD$_2^+$ + HD      &              &              &              &              &              &              & $\checkmark$ &              & $\checkmark$ &              &              \\

\hline
             
        \end{tabular}
      
    \end{table}
    
\end{landscape}
\setcounter{figure}{-1}

\begin{landscape}
    \begin{table}[]
    \tiny
        \centering
          \caption{Dominant formation and destruction processes for
            each molecule in each core.  Reactions are included if
            they contribute more than 10\% to the formation or
            destruction (continued).}\label{tab:reacts2} 
        \begin{tabular}{lccccccccccc}
             Reaction & HOPS108 & Vela C & BHR71 & G300.9 & G304.76 & G309.91 & \multicolumn{2}{c}{Cha2 MMS1} & \multicolumn{2}{c}{Cha2 C2} & TMC-1 \\
             & & & & & & & high n & low n & high n & low n & \\
\hline
\rule{-2pt}{3ex} 
{\bf{CCS (production)}}\\
HC$_2$S$^+$ + e $\rightarrow$ CCS + H      & $\checkmark$ & $\checkmark$ & $\checkmark$ & $\checkmark$ & $\checkmark$ & $\checkmark$ & $\checkmark$ & $\checkmark$ & $\checkmark$ & $\checkmark$ & $\checkmark$\\
HC$_3$S$^+$  + e $\rightarrow$ CCS + CH  & $\checkmark$ & $\checkmark$ & $\checkmark$ & $\checkmark$ & $\checkmark$ & $\checkmark$ & $\checkmark$ & $\checkmark$ & $\checkmark$ & $\checkmark$ & $\checkmark$\\
C$_3$S$^+$ + e $\rightarrow$ CCS + C         & $\checkmark$ & $\checkmark$ & $\checkmark$ & $\checkmark$ & $\checkmark$ & $\checkmark$ &       & $\checkmark$ &      &     & \\
DC$_2$S$^+$ + e $\rightarrow$ CCS + D      &      &      &      &      &      &     & $\checkmark$   & $\checkmark$ & $\checkmark$ &      & \\
CCH + S $\rightarrow$ CCS + H           &   &   &   &     &       &     &       &      &      & $\checkmark$ & \\
DC$_3$S$^+$ + e $\rightarrow$ CCS + CD    &   &    &   &     &      &      &  $\checkmark$ &      & $\checkmark$ &      & \\  
%\\
\rule{-2pt}{3ex} 
{\bf{CCS (destruction)}}\\
CCS + H$_3^+$ $\rightarrow$ H$_2$CS$^+$ + H$_2$ & $\checkmark$ & $\checkmark$ & $\checkmark$ & $\checkmark$ & $\checkmark$ & $\checkmark$& & $\checkmark$ & $\checkmark$ & $\checkmark$ & $\checkmark$\\
CCS + O  $\rightarrow$ CS + CO                  & $\checkmark$ &   & $\checkmark$ & $\checkmark$ &  & $\checkmark$ & & & & $\checkmark$ & \\
CCS + H$^+$ $\rightarrow$ C$_2$S$^+$ + H        &  &  &  & $\checkmark$ &  & & $\checkmark$ & $\checkmark$ & & & \\
CCS + H$_2$D$^+$ $\rightarrow$ HC$_2$S$^+$ + HD & & & & & & & & $\checkmark$ & $\checkmark$ & & $\checkmark$ \\
CCS + HCO$^+$ $\rightarrow$ HC$_2$S$^+$ + CO & & & & & & & & & & & $\checkmark$ \\
CCS + D$_3^+$ $\rightarrow$ DC$_2$S$^+$ + D$_2$ & & & & & & & $\checkmark$ & & $\checkmark$ & & \\
\\
\hline
\rule{-2pt}{3ex} 
{\bf{C$_3$S (production)}}\\
HC$_3$S$^+$ + e $\rightarrow$ C$_3$S + H     & $\checkmark$ & $\checkmark$ & $\checkmark$ & $\checkmark$ & $\checkmark$ & $\checkmark$ & $\checkmark$ & $\checkmark$ & $\checkmark$ & $\checkmark$ & $\checkmark$\\
DC$_3$S$^+$ + e $\rightarrow$ C$_3$S + D      &     &      &      &       &     &      & $\checkmark$ & $\checkmark$ & $\checkmark$ &      & $\checkmark$ \\
HC$_4$S$^+$ + e $\rightarrow$ C$_3$S + CH  &      &       &      &      &      & $\checkmark$ &     &      &      &      & $\checkmark$\\
\rule{-2pt}{3ex} 
{\bf{C$_3$S (destruction)}}\\
C$_3$S + H$_3^+$ $\rightarrow$ HC$_3$S$^+$ + H$_2$          & $\checkmark$ & $\checkmark$ & $\checkmark$ & $\checkmark$ & $\checkmark$ & $\checkmark$ &      & $\checkmark$ & $\checkmark$ & $\checkmark$ & $\checkmark$\\
C$_3$S + H$^+$ $\rightarrow$ C$_3^+$ + H                   &  $\checkmark$ &     & $\checkmark$ & $\checkmark$ & $\checkmark$ & $\checkmark$ & $\checkmark$ &  $\checkmark$  &     &      &  \\
C$_3$S + HCO$^+$ $\rightarrow$ HC$_3$S$^+$ + CO    &       &     & $\checkmark$ &      &      &      &      &       &     & $\checkmark$ & $\checkmark$\\
C$_3$S + C$^+$ $\rightarrow$ C$_4^+$ + S                    &       &     & $\checkmark$ & $\checkmark$ &      &       &      &      &     &      & \\
C$_3$S + C$^+$ $\rightarrow$ C$_3$S$^+$ + C                &       &     & $\checkmark$ & $\checkmark$ &      &       &      &      &     &     & \\
C$_3$S + H$_2$D$^+$ $\rightarrow$ HC$_3$S$^+$ + HD   &        &     &      &     &       &       &      & $\checkmark$ &  $\checkmark$  &     & $\checkmark$ \\
C$_3$S + D$3^+$ $\rightarrow$ DC$_3$S$^+$ + D$_2$       &        &     &      &     &       &       & $\checkmark$ &      & $\checkmark$ &       & \\
\\
\hline
\rule{-2pt}{3ex} 
{\bf{HC$_3$N (production)}}\\
HC$_3$NH$^+$ + e $\rightarrow$ HC$_3$N + H & $\checkmark$ & $\checkmark$ & $\checkmark$ & $\checkmark$ & $\checkmark$ & $\checkmark$ & $\checkmark$ & $\checkmark$ & $\checkmark$ & $\checkmark$ & $\checkmark$\\
CN + C$_2$H$_2$ $\rightarrow$ HC$_3$N + H  & $\checkmark$ & $\checkmark$ & $\checkmark$ & $\checkmark$ & $\checkmark$ & $\checkmark$ & $\checkmark$ & $\checkmark$ & $\checkmark$ & $\checkmark$ & $\checkmark$\\
C$_3$N$^-$ + H $\rightarrow$ HC$_3$N + e   & $\checkmark$ & $\checkmark$ & $\checkmark$ & $\checkmark$ & $\checkmark$ & $\checkmark$ & $\checkmark$& $\checkmark$ & $\checkmark$ & $\checkmark$ & $\checkmark$\\
HC$_3$ND$^+$ + e $\rightarrow$ HC$_3$N + D & & & & & & & $\checkmark$ & \\
DC$_3$NH$^+$ + e $\rightarrow$ HC$_3$N + D & & & & & & & $\checkmark$ & \\
\rule{-2pt}{3ex} 
{\bf{HC$_3$N (destruction)}}\\
HC$_3$N + H$_3^+$ $\rightarrow$ HC$_3$NH$^+$ + H$_2$    & $\checkmark$ & $\checkmark$ & $\checkmark$ & $\checkmark$ & $\checkmark$ & $\checkmark$ & $\checkmark$ &  $\checkmark$ & $\checkmark$& $\checkmark$ & $\checkmark$\\
HC$_3$N + H$^+$ $\rightarrow$ HC$_3$N$^+$ + H        & $\checkmark$ & $\checkmark$ & $\checkmark$ & $\checkmark$ & $\checkmark$ & $\checkmark$ & $\checkmark$ & $\checkmark$ & $\checkmark$ & $\checkmark$ & \\
HC$_3$N + D$_3^+$  $\rightarrow$ DC$_3$ND$^+$ + HD   &     &     &      &      &      &     &   $\checkmark$ &     & $\checkmark$ &      & \\

\rule{-2pt}{3ex} 
{\bf{HC$_5$N (production)}}\\
C$_4$H$_2$ + CN  $\rightarrow$  HC$_5$N + H             & $\checkmark$ & $\checkmark$ & $\checkmark$ & $\checkmark$ & $\checkmark$ & $\checkmark$& $\checkmark$ & $\checkmark$   &   $\checkmark$ & $\checkmark$ & $\checkmark$\\
H$_2$C$_5$N$^+$ + e $\rightarrow$ HC$_5$N + H             & $\checkmark$ & $\checkmark$ & $\checkmark$ & $\checkmark$ & $\checkmark$ & $\checkmark$ & $\checkmark$ & $\checkmark$ &   $\checkmark$  & $\checkmark$ & $\checkmark$ \\
HDC$_5$N$^+$ + e $\rightarrow$ HC$_5$N + D             &     &      &      &      &      &      &  $\checkmark$ &     &  $\checkmark$  &      &     \\
C$_4$HD + CN $\rightarrow$ HC$_5$N + D                &     &      &      &      &      &      &  $\checkmark$ &     &  $\checkmark$  &      &     \\
\rule{-2pt}{3ex} 
{\bf{HC$_5$N (destruction)}}\\
HC$_5$N + H$_3^+$ $\rightarrow$ H$_2$C$_5$N$^+$ + H$_2$          & $\checkmark$ & $\checkmark$ & $\checkmark$ & $\checkmark$ & $\checkmark$ & $\checkmark$ &         & $\checkmark$ & $\checkmark$ & $\checkmark$ & $\checkmark$\\
HC$_5$N + H$^+$ + $\rightarrow$ HC$_5$N$^+$ + H           & $\checkmark$ & $\checkmark$ & $\checkmark$ & $\checkmark$ & $\checkmark$ & $\checkmark$ & $\checkmark$    & $\checkmark$ &       &      &     \\
HC$_5$N + HCO$^+$ $\rightarrow$ H$_2$C$_5$N$^+$ + CO     &     &      & $\checkmark$ & $\checkmark$ &       & $\checkmark$ &   $\checkmark $  &     &  $\checkmark$  & $\checkmark$ & $\checkmark$\\
HC$_5$N + D$_2$H$^+$ $\rightarrow$ HDC$_5$NH$^+$  + HD & & & & & & & & & $\checkmark$ & & \\
             \hline
        \end{tabular}
      
    \end{table}
    
\end{landscape}

\end{appendix}

\end{document}